\documentclass[12pt]{article}
\usepackage{amsmath}
\usepackage{amssymb}
\usepackage{amsxtra}
\usepackage{amsthm}
\usepackage{color}
\usepackage{slashed}

\makeatletter\@addtoreset{equation}{section}\makeatother

\newcommand{\preprint}[1]{\begin{table}[t]  %%
             \begin{flushright}               %%
             {#1}                             %%
             \end{flushright}                 %%
             \end{table}}                     %%
\renewcommand{\title}[1]{\vbox{\center\LARGE{#1}}\vspace{5mm}}
\renewcommand{\author}[1]{\vbox{\center#1}\vspace{5mm}}
\newcommand{\address}[1]{\vbox{\center\em#1}}
\newcommand{\email}[1]{\vbox{\center\tt#1}\vspace{5mm}}

\numberwithin{equation}{section}

\newcommand {\CalN} {\mathcal N}

\newcommand {\BR}   {\mathbb R}
\newcommand {\BZ}   {\mathbb Z}
\newcommand {\BC}   {\mathbb C}

\newcommand {\BQ}   {\mathbb Q}

\newcommand {\BI}   {\mathbb I}

\newcommand {\ve}  {\varepsilon}
\newcommand {\ep}  {\epsilon}

\renewcommand{\Re} {\mathrm{Re}}
\renewcommand{\Im} {\mathrm{Im}}

\newcommand {\p} {\partial}

\DeclareMathOperator{\End}{End}

\DeclareMathOperator{\tr} {Tr}

\DeclareMathOperator{\Pexp} {Pexp}

\newcommand{\SO}{SO}
\newcommand{\Spin}{Spin}

\newcommand{\so}{\mathfrak {so}}

\newcommand{\bea}{\begin{eqnarray}}
\newcommand{\eea}{\end{eqnarray}}

\newcommand{\vv}{\texttt{v}}
\newcommand{\mm}{\texttt{m}}
\newcommand{\ww}{\texttt{w}}

\newcommand{\spt}{\mathrm{spt}}
\newcommand{\scl}{\mathrm{scl}}
\newcommand{\Sph}{S}

\usepackage{hyperref}
\hypersetup{
    colorlinks = true,
    colorlinks = false,
    linkcolor = blue,
    anchorcolor = red,
    citecolor = blue,
    filecolor = red,
    pagecolor = red,
    urlcolor = blue
}

\hypersetup{
    pdftitle =
        Supersymmetric Wilson loop operators
    pdfauthor = 
        Anatoly Dymarsky and Vasily Pestun
    pdfsubject=         
        hep-th,
    pdfkeywords =
        supersymmetry wilson loop operators conformal killing spinor pure spinor AdS/CFT
}

\usepackage[numbers,sort&compress]{natbib}
\usepackage{hypernat}

\usepackage{graphicx}

\begin{document}

\unitlength = .8mm

\bibliographystyle{utphys}

\begin{titlepage}
\begin{center}
\hfill \\
\hfill \\

\preprint{SITP-TH-09/33
\\ ITEP-TH-27/09}

\title{Supersymmetric Wilson loops in $\CalN=4$ SYM \\ and pure spinors}

\renewcommand{\thefootnote}{\fnsymbol{footnote}}
\author{Anatoly Dymarsky$^{a,\,*}$ and
Vasily Pestun$^{b,\,}$\footnotemark}
\footnotetext{On leave of absence from ITEP, 117218, Moscow, Russia}

\address{$^a$  School of Natural Sciences, Institute for Advanced Study, \\
Princeton, NJ, 08540 USA}
\email{dymarsky@ias.edu}

\address{$^b$ Center for the Fundamental Laws of Nature
\\Jefferson Physical Laboratory, Harvard University,\\
Cambridge, MA 02138 USA}
\email{pestun@physics.harvard.edu}

\end{center}

\abstract{
We study supersymmetric Wilson loop operators in four-dimensional $\CalN=4$ super Yang-Mills theory. 
%We identify all supercharges $Q$  that admit a supersymmetric
% Wilson loop operator $W$. 
We show that the contour of a supersymmetric Wilson loop is either an orbit of some conformal transformation of the space-time (case I), 
or an arbitrary contour in the subspace where local superalgebra generator is a pure spinor (case II). 
 In the more interesting case II we find and classify all pairs $(Q,W)$ of the supercharges and the corresponding operators 
 modulo the action of the global symmetry group. 
 }

\vfill

\end{titlepage}

\eject

\tableofcontents

\section{Introduction}

The four-dimensional maximally supersymmetric Yang-Mills theory ($\CalN=4$ SYM) is a fascinating 
model which exhibits rich but rigid mathematical structure. Thanks to
the AdS/CFT correspondence
\cite{Polyakov:1987ez,Maldacena:1997re,Witten:1998qj,Gubser:1998bc} the
theory has been in focus of theoretical research for the past
decade. Many interesting results including those based on integrability
\cite{Beisert:2006ez,Bern:2006ew} suggest that $\CalN=4$ SYM may have an
exact solution in the large $N$ limit at least in the supersymmetric
sector. This motivates our interest in studying the supersymmetric
sector to identify non-local gauge invariant observables.

The $\CalN=4$ SYM  
is a superconformal theory. The fermionic subspace 
of its superconformal algebra is generated by Poincare supercharges $Q_\alpha$
and special conformal supercharges $S^{\alpha}$. In the scope of the present work 
we call an operator supersymmetric if there exists 
at least one non-zero linear combination of $Q_{\alpha}$ and
$S^{\alpha}$ that annihilates the operator.

In this paper we are interested in one-dimensional non-local operators.
Familiar examples of such operators are 't
Hooft and Wilson loop operators. Presently we focus on supersymmetric 
Wilson loop operators, which are obtained from the
ordinary Wilson loops by coupling them to the scalars of the $\CalN=4$ SYM \cite{Maldacena:1998im}.
We consider the theory on the Euclidean space-time $\BR^4_{\spt}$.

A number of such supersymmetric Wilson loops  have been found and analyzed previously, see e.g.
\cite{Drukker:2000rr,Erickson:2000af,Semenoff:2001xp,Erickson:1999uc,
  Drukker:2007qr,Drukker:2009sf,Giombi:2009ds, Bassetto:2009rt,Zarembo:2002an}. 
All supersymmetric Wilson loops that have been studied previously are captured 
by two classes:  the loops of arbitrary shape on $\BR^4_{\spt}$ found by Zarembo in
\cite{Zarembo:2002an} and the loops of arbitrary shape on a three-sphere
$S^3 \subset \BR^4_{\spt}$ found by
Drukker-Giombi-Ricci-Trancanelli (DGRT) in \cite{Drukker:2007qr}. 
Zarembo's loops on $\BR^4_{\spt}$ are the same Wilson loops which appear in topological
Langlands twist of $\CalN=4$ SYM \cite{Kapustin:2006pk}; they have
trivial expectation value. The string dual surfaces to these loops were described in \cite{Dymarsky:2006ve}. 
The most familiar example of the loops in DGRT class is the 1/2 BPS circular loop coupled to one
of the scalars; this Wilson loop can be computed exactly by Gaussian matrix model
\cite{Erickson:2000af,Drukker:2000rr,Pestun:2007rz} and the results
agree with the string dual computation. 
The subset of DGRT loops restricted to $S^2$ was also recently studied
in great details and a connection between this sector
of $\CalN=4$ SYM and two-dimensional Yang-Mill on $S^2$ was established
\cite{Drukker:2007yx,Drukker:2007dw,Young:2008ed,Bassetto:2008yf,Pestun:2009nn,Bassetto:2009rt,Giombi:2009ds,Giombi:2009ek,Giombi:2009ms}. 
% These Wilson loops share some common features. 
%The contour $\gamma(s)$ is restricted to a subspace $\Sigma\subset
%\BR^4$, where $\Sigma$ is either $\BR^4$ or $S^3$ but is arbitrary
%otherwise and the scalar couplings $v^A$ are uniquely defined by the
%shape of the contour. Moreover the scalar couplings $v^A$ are pure
%imaginary. 
It has not been clear whether these two classes capture all possible
supersymmetric Wilson loops. 
%(\ref{eq:extended-Wilson-loop-definition}). 

In this note we give a systematic answer to this question. 
We find all possible  Wilson loop
 operators $W$ 
% (\ref{eq:extended-Wilson-loop-definition})% 
that are invariant at least under one superconformal symmetry $Q$. 
Moreover, we classify the interesting subclass of pairs $(Q,W)$ modulo 
equivalence under the action of the superconformal group of the $\CalN=4$ SYM.

%Our goal would be to classify all distinct pairs of Wilson loops and the corresponding supercharges $Q$ modu%lo the action of the global symmetry group of the theory $\SO(5,1)\times \SO(6)$. 

We find new supersymmetric Wilson loops
which has not been identified before. In many cases the new operators
involve complex couplings to the scalar fields that
 clearly distinguishes them from the previously studied cases. 
In certain cases the new operators could be related to the previously
known ones
 by a complexified conformal transformation.   
However,
unless we define the theory on the complexified
space-time, and stay in the framework of the conventional theory formulated in the real Euclidean
space, the novel operators are not equivalent to the known ones. 

The crucial ingredient in our construction are the ten-dimensional pure spinors. 
Their relevance is not so surprising given that the
four-dimensional $\CalN=4$ SYM is a dimensional reduction of the
ten-dimensional $\CalN=1$ SYM, where pure spinors appear 
naturally \cite{Berkovits:2001rb,Baulieu:2007ew,Berkovits:1993zz}.
The space-time dependent spinor $\ve$ that parametrizes the superconformal transformations of
$\CalN=4$ SYM, can be viewed as a reduction of a chiral ten-dimensional
spinor.

% While checking whether a given Wilson loop operator coupled to
% scalars is supersymmetric at a particular point $x$ of the space-time 
% with respect to
% local supersymmetry generated by $\ve(x)$, one gets a certain 
% system of equations on the scalar couplings and the direction of the
% curve at $x$. 

Locally, at a point $x$ of the space-time, Wilson loop operator can be
locally described
 by the tangent vector to the curve and the scalar couplings at $x$. We combine this data
into ten-dimenensional vector $v(x)$. 
If we want to find supersymmetric Wilson loops with respect
to a supersymmetry generated by a given spinor $\ve(x)$, we get a certain
system of equations on $v(x)$. This system of equations might
be of different types depending on $\ve(x)$. 
If $\ve(x)$ is not a pure spinor, then the system has the
unique solution, so that the tangent to the curve and the scalar
couplings at $x$ are completely fixed. Namely, the tangent to the
curve and the scalar couplings could be combined into a ten-dimensional
vector $v(x)$. This vector, projectively, is precisely the ten-dimensional vector
constructed in the canonical way as the bilinear in $\ve(x)$.
The curves, resulting in this way from a generic 
supersymmetry parameter $\ve(x)$, are nothing else but the orbits
of the conformal transformation generated by $Q_{\ve}^2$. If we ask 
for the orbits to be compact, then modulo
conformal equivalence, 
%the $Q^2$ sits inside the Lorents subgroup $\SO(4)$ of
%the conformal group. Then 
the only resulting compact curves are the $(p,q)$ 
Lissajous figures where $\frac p q \in \BQ$ is the rational ratio of two eigenvalues
of the $\so(4)$ matrix which represents the action of $Q^2$ on the
space-time $\BR^4_{\spt}$. 
  
If $\ve(x)$ is pure then there are more solutions 
for the vector $v(x)$ (which  tangent to the curve at $x$ and scalar couplings described
together by the ten-dimensional vector $v(x)$.  
More precisely, a pure spinor $\ve(x)$ defines ten-dimensional
almost complex structure $J(x)$, and then the supersymmetry condition of the
Wilson loop at $x$ translates to the condition that $v(x)$ is
anti-holomorphic vector with respect to $J(x)$. 
On the subspace $\Sigma$ of the space-time where $\ve(x)$ is pure there is
richer space of solutions for supersymmetric Wilson loops. Generically, 
for any curve sitting inside $\Sigma$ one can find scalar couplings to
make supersymmetric Wilson loop.

% For (\ref{eq:extended-Wilson-loop-definition}) to  be supersymmetric the argument of the exponent 
% $\int A_M v^M ds$ should be invariant under the supersymmetry transformation $\delta A_M=\ve \Gamma_M \Psi$
% at each point of the contour $\gamma$. The chiral conformal Killing spinor $\ve(x)$ 
% that parametrizes the supersymmetry transformation is a function of two constant spinors of opposite chirality
% \bea
% \label{eq:KillSp}
% \ve(x)=\ve_s+x^\mu \Gamma_\mu \ve_c\ .
% \eea  
% For $A_M v^M$ to be invariant we must require 
% \bea
% \label{eq:VE}
% v^M\Gamma_M \ve=0\ ,
% \eea
% at each point on the contour.
% The equation (\ref{eq:VE}) immediately leads us to pure spinors. For a general $\ve$ there is only one solution $v^M\cong\ve \Gamma^M \ve$. It completely specifies the Wilson line upon given $(\ve_s,\ve_c)$ and starting point $x^\mu(s=0)$.  Nevertheless if 
% $\ve$ is pure i.e. when $\ve \Gamma^M \ve$ vanishes, $\ve$ defines a five-dimensional space 
% of ``anti-holomorphic'' vectors $v^M$ that solve (\ref{eq:VE}). Therefore to construct a variety of  Wilson loop operators 
% we should find a subspace $\Sigma\subset \BR^4$ where $\ve(x)$ is pure for the given $(\ve_s,\ve_c)$. Then at each point $x\in \Sigma$ 
% one can choose any   anti-holomorphic $v^M$. As a result we have a rich space of Wilson loop operators
% parametrized not only by the contour $x^\mu(s)\subset \Sigma$ but also by some functional parameters governing the scalar couplings $v^A$.   

The supersymmetry spinor $\ve(x)$ of $\CalN=4$ SYM can 
be extended to the $AdS_5\times S^5$ space where it plays the role of
the supersymmetry spinor of the IIB String Theory. 
Similarly the space $\Sigma$ where $\ve(x)$ is pure can be extended to
the the subspace $\Sigma_{\BC}$ in $AdS_5\times S^5$. 
The pure spinor defines an almost complex structure $J
\in \End(T_{\Sigma_{\BC}} + N_{\Sigma_{\BC}})$, 
where $T$ and $N$ stand respectively for the tangent and normal bundles of
$\Sigma_{\BC} \subset AdS_5 \times S^5$. 

We conjecture that for a Wilson loop operator with the contour in $\Sigma$ the classical dual string worldsheet lies 
on $\Sigma_{\BC}$ and is pseudo-holomorphic with respect to $J$. This is supported by the fact that the
$J$-pseudo-holomorphic solution is necessarily supersymmetric.  
Thus the results \cite{Dymarsky:2006ve,Drukker:2007qr} developed earlier
for the string duals of Zarembo loops \cite{Zarembo:2002an} and DGRT loops
\cite{Drukker:2007qr}
are the particular examples of this general picture.

The structure of the paper is as follows.
In section \ref{se:conventions} we summarize our conventions on
$\CalN=4$ SYM and superconformal transformations in Euclidean
space-time. 
In section \ref{se:susy-loops-pure-spinors} 
we give general construction of supersymmetric Wilson operators and
relate that to pure spinors. In section
\ref{se:pure-spinor-manifold-sigma} we find the pure-spinor-surfaces
$\Sigma$  construct the  supersymmetric Wilson loop operators. The next section
\ref{se:classification-loops} deals with
classification of the pairs $(Q,W)$ related to pure spinors modulo 
equivalence under the action of the superconformal group of the $\CalN=4$ SYM.

\section{Conventions\label{se:conventions}}
%\subsubsection{The fields and symmetries}
We consider the Euclidean space-time $\BR^{4}_{\spt}$ equipped with the standard flat unit
metric. 

We take the action of the $\CalN=4$ SYM gauge theory with gauge
group $G$  on $\BR^4_{\spt}$  to be
\begin{equation}
  \label{eq:N4-Lagragian-d-4}
\begin{split}
  S = -\frac {1} {2 g_{YM}^2} \int d^4 x \tr \left(\frac 1 2 \left(F_{\mu \nu} F^{\mu \nu}
+ D_{\mu} \Phi_A D^{\mu} \Phi^{A} + \frac 1 2 [\Phi_A \Phi_B] [\Phi^A \Phi^B]\right) -
\right. \\ \left. - \Psi \Gamma^{\mu} D_{\mu} \Psi - \Psi \Gamma^{A} [\Phi_A \Psi]  \right)\ .
\end{split}
\end{equation}

The indexes $\mu, \nu = 1,\dots,4$ label the directions in the
space-time, the indices $A,B=5 \dots 10$ label the directions
in the target space of scalars. We often combine indexes $\mu,\nu$ with $A,B$
into ten-dimensional indexes $N,M=1\dots 10$, and the gauge
field $A_{\mu}$ and the scalar fields $\Phi_{A}$ into 
$ A_M :=(A_\mu,\Phi_A)$. That could be interpreted about as dimensional
reduction of the gauge field of $d=10$ $\CalN=1$ SYM. All fields take
value in the Lie algebra of the gauge group $G$, the conventions 
for the covariant derivative and the curvature are $D_{\mu} = \p_{\mu} +
A_{\mu}$ and $F_{\mu\nu} = [D_{\mu}, D_{\nu}]$.

The fermionic fields $\Psi$ are sixteen-component spinors
obtained by dimensional reduction from the chiral spin representation of
$\Spin(10,\BR)$ which we call $S^{+}$. 
The chiral spin representation of $\Spin(10,\BR)$  dual to $S^{+}$ is called $S^{-}$.  
The matrices $\Gamma^M: S^{+} \to S^{-}$ are the $16\times 16$ matrices which are the 
chiral blocks of the $32 \times 32$ ten-dimensional Dirac
gamma-matrices $\gamma_{\mu}$. We use conventions where 
\begin{equation}
  \label{eq:32gamma-matrice-16-gamma-matrices}
\gamma_{M}:
\begin{pmatrix}
  S^{+} \\
  S^{-}
\end{pmatrix} \to
\begin{pmatrix}
  S^{+} \\
  S^{-}
\end{pmatrix}, \quad 
  \gamma_{M} =
  \begin{pmatrix}
    0 &  \Gamma_{M}^{*}\\
    \Gamma_{M} & 0
  \end{pmatrix}, \quad \Gamma_{M} = \Gamma_{M}^T, \quad \{\gamma_{M},
  \gamma_{N}\} = \delta_{MN}.
\end{equation}

 The explicit form of $\Gamma^M$ can be found in Appendix A. 
In ten dimensions there is no
need for complex or Dirac-like conjugation to write down a fermionic bilinear like $\Psi \Gamma^{M}  \Psi$ 
which is literally  
\begin{equation}
  \label{eq:fermionic-term-explicit}
  \sum_{\alpha, \beta =1}^{16} \Psi^{\alpha} \Gamma^{M}_{\alpha \beta} \Psi^{\beta}\ .
\end{equation}
We use the indexes $\alpha, \beta = 1 \dots 16$ to denote
the sixteen components of $S^{+}$ spinors such as $\Psi_{\alpha}$. 
Since we consider the theory as dimensionally reduced from Euclidean
space $\BR^{10}$ rather than Minkowski space $\BR^{9,1}$, we do not
require $\Psi$ to be real. However, in the path integral we integrate only over $\Psi$ but not over their
complex conjugates. This is consistent because complex conjugate to $\Psi$
never appears in the action or anywhere else.

We consider the superconformal
transformations
\begin{equation}
  \label{eq:super-conformal-transformations}
\begin{aligned}
\delta A_{M} &= \ve \Gamma_{M} \Psi, \\  
\delta \Psi  &=  \frac 1 2   F_{MN} \Gamma^{MN} \Psi +  \Phi_{A} \Gamma^{\mu A}  \nabla_{\mu} \ve,
\end{aligned}
\end{equation}
where spinor $\ve(x)$ is a parameter. 
We treat the spinor $\ve(x)$ as a bosonic parameter of the fermionic supersymmmetry
transformation. It transforms in the same spin representation as $\Psi$,
i.e. in $S^{+}$. 
The $\CalN=4$ SYM action (\ref{eq:N4-Lagragian-d-4}) is invariant under
(\ref{eq:super-conformal-transformations}) if $\ve(x)$ is a conformal
Killing spinor (twistor spinor) \cite{MR944085}.

By definition, a conformal Killing spinor $\ve(x)$ is a solution of 
the twistor equation (see \cite{MR1164864,MR1131908,MR1164864} for a review on
conformal Killing spinors)
\begin{equation}
  \label{eq:conformal-Killing-spinor}
  D_{\mu} \ve =  \frac 1 4 \Gamma_{\mu} \slashed{D}\ve. 
\end{equation}
We use the notation $\tilde \ve = \frac 1 4 \slashed{D} \ve $, so the
conformal Killing spinor equation is $D_{\mu} \ve = \Gamma_{\mu} \tilde \ve$.
The solutions on $\BR^{4}$ are parametrized 
 by two constant spinors,  which we call $\ve_s \in S^{+}$ and $\ve_c \in S^{-}$
\begin{equation}
  \label{eq:conformal-killing-solutions}
  \ve(x) = \ve_{s} + x^{\mu} \Gamma_{\mu} \ve_{c}\ .
\end{equation}
In total there are $16+16 = 32$ complex generators of superconformal symmetries. 
The spinor $\ve_{s}$ generates the usual supersymmetries associated with 16 supercharges which are
customarily called $Q_{\alpha}$, and the spinor $\ve_{c}$ 
generates the remaining special conformal supersymmetries associated with 16 supercharges which are customarily
called $S^{\alpha}$. The supersymmetry transformation $Q_{\ve}$ is given
by $Q_{\ve} = \ve^{\alpha}_s Q_{\alpha} + \ve^{c}_{\beta} S^{\beta}$.

\section{Supersymmetric Wilson loops and pure spinors \label{se:susy-loops-pure-spinors}}
For a closed contour $\gamma: S^{1} \to \BR^{4}_{\spt}$ and a representation $R$ of the gauge group, 
a Wilson loop operator $W_R(\gamma)$ is the trace in the representation $R$ of the path ordered 
integral of the gauge field along the contour $\gamma$ 
\begin{equation}
  \label{eq:Wilson-loop-definition}
  W_{R}(\gamma) = \tr_{R} \Pexp \oint_{\gamma} A_{\mu} dx^{\mu}.
\end{equation}
%This operator is, of course, gauge invariant but it is not supersymmetric. 
A natural generalization of  (\ref{eq:Wilson-loop-definition}) for a theory with 
adjoint scalars is obtained by coupling to the scalar fields $\Phi_A$ \cite{Maldacena:1998im,Rey:1998ik}
\begin{equation}
\begin{aligned}
\label{eq:extended-Wilson-loop-definition}
  W_{R}(\hat \gamma) = \tr_{R} \Pexp \oint_{\gamma}  A_{M} v^{M} ds  =  \\
                     = \tr_{R} \Pexp \oint_{\gamma} ( A_{\mu} v^{\mu} + \Phi_{A} v^{A}) ds\ ,
\end{aligned}
\end{equation}
where the generalized contour 
$\hat \gamma = (x^{\mu}(s), v^{A}(s))$ 
is now defined by specifying the four-dimensional tangent velocity vector $v^{\mu}(s)=dx^\mu/ds$ and a six-vector 
of scalar couplings $v^{A}(s)$. To make the usual sense of the contour in the real space-time
the tangent vector $v^\mu(s)$ must be real. At the same time the scalar couplings $v^{A}(s)$ generically 
could be complex. Our notation $v^{A}$ is related to the common notation
$\theta^{A}$, used in the literature on the
subject~\cite{Maldacena:1998im,Zarembo:2002an,Drukker:2007qr},
via  $v^{A} =
i \theta^A$.  A local operator, say $\frac {1}{J!} \tr_R (\Phi_5 + i
\Phi_6)^J$ is  also captured by the generic definition
(\ref{eq:extended-Wilson-loop-definition}). 
It corresponds to contour $\gamma$ which is point in $\BR^4_{\spt}$
but a unit interval in $\BR^6_{\scl}$ such that $\int_{\gamma} v^{A} =
(1,i,0,0,0,0)$ and taking $J$-th term in Taylor series of the exponent
expansion.

For the Wilson loop (\ref{eq:extended-Wilson-loop-definition}) to be supersymetric 
  $v^M A_M$  must be invariant under
  (\ref{eq:super-conformal-transformations}), that is
\begin{equation}
  \label{eq: supersymmetric-condition}
  v^{M}(x) \ve(x) \Gamma_{M} \Psi = 0 \quad 
\end{equation}
has to vanish for any $\Psi$ at each point on the contour $\gamma(s)$. This implies 
\begin{equation}
  \label{eq:susy-loop-condition}
  v^{M}(x) \Gamma_{M} \ve(x) = 0\ .
\end{equation}

To find all possible solutions of (\ref{eq:susy-loop-condition}), we first 
consider the problem locally a point $x$. We assume that we have a generic spinor $\ve$,
and we want to identify the space of
directions $L \subset \BR^{10} \otimes \BC$ which annihilate $\ve$ under the Clifford
action: 
\begin{equation}
  \label{eq:annih-equation}
  v^M \Gamma_M \ve =  0,\quad v \in L.
\end{equation}
At this moment we allow $v^M$ to be complex and consider possible reality
conditions later.

For any $\ve$ we can canonically construct a bilinear vector
$u^{M}(\ve)$ as
\begin{equation}
  \label{eq:bilinear}
  u^{M} = \ve \Gamma^{M} \ve.
\end{equation}
Now, depending on whether $u^M  = 0$ or $u^M \neq 0$ we have two distinct cases. 
If $u^M = 0$ the spinor $\ve$ is \emph{pure spinor}, and if $u^M \neq 0$ the
spinor $\ve$ is \emph{not pure spinor}. This requires some clarification which will be given below.

First consider the generic case when $u^M(x) \neq 0$ ($\ve$ is not pure spinor).
It is a simple exercise to
show that in ten dimensions $v^M = \lambda u^M, \lambda \in \BC$ is the only solution to (\ref{eq:annih-equation}) unless $u^M =0$.
The fact that $v^{M} = u^{M}$ is a solution follows from the following identity for the ten-dimensional gamma-matrices
\begin{equation}
  \label{eq:triple-identity}
  \Gamma^M_{\alpha (\beta} \Gamma^M_{\gamma \delta)}  = 0 \ .
\end{equation}
(This Fierz identity is used to establish supersymmetry of $d=10$
$\CalN=1$ SYM.) The proof of the uniqueness of the solution $v^M \sim u^M$ for $u^M\neq 0$ can
be found in Appendix \ref{se:unique-sol}.

Second consider the case when $u^M(x) = 0$ ($\ve$ is pure spinor). In the ten-dimensional space the equation 
\begin{equation}
\label{eq:pure-condition}
\ve \Gamma^{M} \ve = 0 
\end{equation}
is equivalent to saying that $\ve$ is a \emph{pure spinor}
\cite{Berkovits:2001rb}. 
Generically, a spinor $\ve$ for $\Spin(\BR^{2n})$ is called \emph{pure}
if it is annihilated by half of gamma-matrices: there exists a half-dimensional subspace $L \subset  \BR^{2n} \otimes \BC$ 
such that 
\begin{equation}
  \label{eq:pure-definition}
  v^{M} \Gamma_{M} \ve = 0  \Leftrightarrow v \in L\ .
\end{equation}
A pure spinor $\ve$ defines a complex structure on the vector space
$\BR^{2n} \otimes \BC$ by saying that $L$ is
the space of anti-holomorphic vectors $L = V^{(0,1)}$. 
In general, a complex structure on vector space $\BR^{2n}$ can be
defined as a $2n \times 2n$ antisymmetric matrix $J$ such that $J^2 =
-1$. 
Under action by $J$, the complexified vector space $\BR^{2n} \otimes
\BC$ splits as $\BR^{2n}\otimes \BC = V^{(1,0)} + V^{(0,1)}$, 
where holomorphic $V^{(1,0)}$ is the $+i$-eigenspace of $J$ and 
anti-holomorphic $V^{(0,1)}$ is the $-i$-eigenspace of $J$. 

Therefore, whenever $\ve \Gamma^{M} \ve = 0$, the solutions to the local
supersymmetry equation (\ref{eq:annih-equation})
are the  \emph{anti-holomorphic} vectors $v^M$ with respect to the
complex structure $J_{\ve}$. 
 In our case $V^{(0,1)}_{\ve}$ is a five-dimensional complex vector space.

Now we can return back to the Wilson loop (\ref{eq:extended-Wilson-loop-definition}) and describe the operators invariant under a superconformal generator  $Q_{\ve}$.  
At a generic point in the space-time $x$ where $u^{M}(\ve(x)) \neq 0$ , locally, the
only supersymmetric  Wilson loop is 
\begin{equation}
  \label{eq:generic-point}
   \Pexp \int_{\gamma} (A_{\mu} u^{\mu} + \Phi_{A} u^{A}) \frac {ds} {(u^{\mu} u_{\mu})^{1/2}}\ .
\end{equation}
The tangent to the contour $\gamma$, specified by  $x^{\mu}(s)$, must be
aligned with  $u^{\mu}(x)$. 
In order for the contour $\gamma$ to be in  $\BR^{4}_{\spt}$  the vector
$u^{\mu}$ must be projectively real, i.e. there is $\lambda \in \BC^{*}$
such that $\frac {dx^{\mu}}{ds} = \lambda u^{\mu}$ is real. 

The vector field $u^{\mu}(\ve(x))$  has simple geometrical
interpretation. It is the vector field of the infinitesimal conformal
transformation generated by $Q_{(\ve_s,\ve_c)}^2$. One can check (see e.g. \cite{Pestun:2007rz}) that
the action of $Q_{\ve}^2$ on any field $\phi$ of the theory is represented as 
\begin{equation}
  Q_{\ve}^2 \phi(x) = (-L_{u} - G_{u^M A_M}  - R - \Omega)
  \phi(x) \,
\end{equation}
where $L_{u}$ is the Lie derivative in the direction of $u$,
 the symbol $G_{u^{M}A_{M}}$ denotes  gauge transformation, the symbol $R$  is the
 $R$-symmetry transformation and the symbol $\Omega$ is a local scale transformations
 acting on fields according to their conformal dimensions.

In  points $x$ where $u^\mu(x) = 0$ but $u^M \neq 0$  the supersymmetric Wilson loop reduces to a local operator
\begin{equation}
  \label{eq:gener-Wilson-local}
  W_{R}(x^{\mu},u) = \tr_{R} \exp (  \lambda u^{A}
  \Phi_{A}(x^\mu)),\quad   \lambda \in \BC^{*}.
\end{equation}

The most interesting case is when $u^M$ vanishes on some subspace $\Sigma_{\ve} \subset \BR^{4}_{\spt}$ 
\begin{equation}
  \label{eq:FixU-definition}
  \Sigma_{\ve} = \{ x \in \BR^4_{\spt} |  u^M (\ve(x)) = 0 \}\ .
\end{equation}
The spinor $\ve(x)$ is pure everywhere on $\Sigma_{\ve}$. Locally at a given point $x$, 
the tangent $v^\mu$ and the scalar couplings $v^A$ of supersymmetric
Wilson loop must be components of an anti-holomorphic vector $v^M \in V^{(0,1)}_{\ve(x)}$.  

To find all  Wilson loop operators in this class, for each $Q_{\ve}$
we find pure-spinor-surface  $\Sigma_{\ve}$  and the bundle of anti-holomorphic vectors $V^{(0,1)}_{\ve} \to
\Sigma_{\ve}$. For each contour $\gamma$ such that the tangent vector
$v^{\mu}$
is a projection to $T_{\Sigma_{\ve}}$ of some section $v$ of $V^{(0,1)}$
we can associate supersymmetric Wilson loop.  

We remark that we do not require any integrability condition
 for the almost complex structure along $\Sigma$ as it was not 
needed to establish supersymmetry of the Wilson loop operators.
Unless explicitly stated otherwise, by complex structure 
we always mean an almost complex structure.

\subsection{\label{se:pure-spinors-in-bulk} Pure spinors in $AdS_5\times S^5$}
The conformal Killing spinor (\ref{eq:conformal-killing-solutions}) 
can be extended from the boundary $\BR^{4}_{\spt}$ of $AdS_5$  into the bulk of $AdS_5\times S^5$ 
\begin{equation}
\label{eq:Killing-spinor-bulk}
\ve_{AdS}(x^M)={1\over \sqrt{z}}\left(\ve_s+x^M\Gamma_M \ve_c\right),
\end{equation}
where it becomes the supersymmetry transformation parameter for the
theory in the bulk, see e.g. \cite{Claus:1998yw,Drukker:2007qr}. 
The explicit formula (\ref{eq:Killing-spinor-bulk}) is presented in the
vielbein for the spin bundle over $AdS_5 \times S^5$ associated
canonically to the coordinates $(x^{\mu}, y^{A})$ on $AdS_5 \times S^5$
in which metric has the form 
\begin{equation}
  \label{eq:metric-ads-Y}
   ds^2 =   y^2 dx^{\mu} dx_{\mu} + \frac{ dy^A dy_A} {y^2}.
\end{equation}
The coordinates  $y^{A}$ are related to coordinates $z^A$ as $y^{A} =
z^{A}/z^2$ where in coordinates $x^{M} = (x^{\mu}, z^{A})$ the same
  metric (\ref{eq:metric-ads-Y}) is 
\begin{equation}
\label{eq:metric-ads-X}
   ds^2=  G_{MN}dx^M dx^N={dx^\mu dx_\mu+dz^A dz_A\over z^2}.
\end{equation}

The subspace $\Sigma\subset\BR^4_{\spt}$ where $\ve(x)$ is pure can also
be extended to $\Sigma_{\BC}\subset AdS_5\times S^5$. 
 Then $\ve_{AdS}$ defines an almost complex structure
 $J$ on $\Sigma_{\BC}$, more precisely $J$ is a section of $\End(T_{\Sigma_{\BC}} +
N_{\Sigma_{\BC}})$ such that it is compatible with the metric and that
$J^2 = -1$.
 We conjecture that the classical stringy world-sheet dual to the supersymmetric Wilson loop operator with contour living on $\Sigma$ will be given by a pseudo-holomorphic surface in $\Sigma_{\BC}$. In support of this idea we show that such a solution would satisfy the $\kappa$-symmetry condition in the bulk i.e. will be supersymmetric.

We choose the coordinates on the stringy world-sheet such that the induced metric is flat $g_{\alpha\beta}=\delta_{\alpha\beta}$. In this notations the pseudo-anti-holomorphic surface is given by
\begin{equation}
\label{eq:pseudo-holomorphic}
V_\alpha^M=\partial_\alpha X^M-\epsilon_{\alpha\beta}J^{M}_N\partial_\beta X^N=0\ .
\end{equation}
This condition guarantees that the corresponding profile is
supersymmetric i.e. it satisfies the $\kappa$-symmetry condition
\bea
\label{eq:kappa-symmetry}
(\epsilon_{\alpha\beta}\partial_\alpha X^M \partial_\beta X^N \Gamma_{MN}-i\delta_{\alpha\beta}\partial_\alpha X^M \partial_\beta X^N G_{MN})\ve_{AdS_5}=0\ .
\eea 
Following 
\cite{Drukker:2007qr} we prove 
 (\ref{eq:kappa-symmetry}) by showing that 
\bea
\label{eq:phol}
\partial_\alpha X^M(\delta^{N}_M+iJ^{N}_M)\Gamma_N \ve_{AdS}=0\ ,
\eea 
is satisfied (the $\kappa$-symmetry condition can by obtained from (\ref{eq:phol}) by multiplying it by $\partial_\alpha X^M\Gamma_M$). The latter is obvious because the vector $\partial_\alpha  X^M(\delta^{N}_M+iJ^{N}_M)$ is anti-holomorphic i.e. it is an $-i$-eigenvalue of the pseudo-complex structure $J$. Therefore
it annihilates the spinor $\ve_{AdS_5}$ according to the {\it definition} of $J$. 

We remark that this result for the specific cases of the
strings dual to  Zarembo's loops \cite{Zarembo:2002an} and
DGRT's loops \cite{Drukker:2007qr} was obtained  in
\cite{Dymarsky:2006ve,Drukker:2007qr}. However, there the
pseudo-holomorphic structure $J$ appeared as an extra input, not directly
related to $\ve$, and (\ref{eq:phol}) was established with help of 
the explicit form of $J$ and $\ve_{AdS}$. 
We cosntruct $J$ canonically starting from an arbitrary superconformal
symmetry parameter $\ve$ at points where $\ve$ is pure.

In addition, one can easily see that the supersymmetry implies that the world-sheet is
psedo-holomorphic provided that it lies in $\Sigma_{\BC}$. To show that one can multiply (\ref{eq:kappa-symmetry})
by $\ve_{AdS_5}$ from the right and use that $(V_\alpha^N)^2=0$ implies
$V_\alpha^N=0$.  We do not have a general argument why the
 world-sheet dual to Wilson loop in $\Sigma$ must sit inside $\Sigma_{\BC}$, but that seems to be a
reasonable conjecture. 

The pseudo-holomorphic surface is always calibrated by some calibration form $P[J]$ as follows from  the following inequality
\begin{equation}
\int d^2\sigma G_{MN}V_\alpha^M V_\alpha^{N} \geq 0,
\end{equation}
and hence
\begin{equation}
S_{string}\geq \int P[J], \quad  J_{[MN]}=G_{ML}J^{L}_N.
\end{equation}
In general the calibration form $J$ is not closed, therefore
we cannot immediately compute the classical action as a functional of the boundary 
conditions.
%
%Therefore we cannot immediately compute the action and hence the
%expectation value of the corresponding Wilson loop operator at 
%strong coupling without actually finding the profile.

\section{Pure-spinor surfaces $\Sigma$\label{se:pure-spinor-manifold-sigma}}

In this section we will find explicitly all superconformal generators
$Q_{\ve}$ that admit a non-trivial pure-spinor-surface $\Sigma_{\ve}$. 
We call $\Sigma_{\ve}$ non-trivial if it has at least one component of
positive dimension.  

%other words, we are not interested in the case when $\Sigma$ is a set of separated points, since there are no non-local Wil%son loop operators associated with $\Sigma$ in this 
%case. 
We pick
any connected component of positive dimension 
of $\Sigma_\ve$ and call it $\Sigma_{\ve}$ in what follows.\footnote{Actually we will
see later that $\Sigma_{\ve}$ is always connected.} 

We choose any point in $\Sigma_{\ve}$ to be an origin of the
coordinate system in $\BR^4_{\spt}$. 
In this coordinate system the conformal Killing
spinor $\ve$ has the form 
\begin{equation}
  \label{eq:conformal-Killing-new-system}
  \ve(x) = \ve_{s}  + x^{\mu} \Gamma_{\mu} \ve_{c}\ ,
\end{equation}
where $\ve_{s} = \ve|_{x=0}$ is pure. Our goal is to find  for which 
$\ve_{c}$ there is a nontrivial pure spinor surface
$\Sigma_{\ve}$ (\ref{eq:FixU-definition}) and what shape $\Sigma_{\ve}$ has. 
From the definition of $\Sigma_{\ve}$ and (\ref{eq:pure-condition}) it follows 
that $\Sigma$ is an intersection of 10 quadric hypersurfaces in $\BR^{4}_\spt$. 
Potentially $\Sigma$ can have a complicated shape. 
It turns out that it is easier first to solve a more generic problem in ten dimensions.
For that reason we formally continue the
 conformal Killing spinor (\ref{eq:conformal-Killing-new-system}) from $\BR^{4}_\spt$ to $\BR^{10}$ by
 replacing $x^{\mu} \Gamma_{\mu}$ by $x^{M} \Gamma_{M}$. We have seen in the previous section
 that the extended spinor $\ve(x)$  in ten dimensions (\ref{eq:Killing-spinor-bulk}) plays the role of 
the supersymmetry parameter of string theory in $AdS_{5} \times S^{5}$. 

\subsection{\label{se:form-notations-pure-spinor-constraints} Form notations and pure spinor constraints}

We start by introducing the subsurface $\Sigma_{\BC} \subset \BR^{10}$  where the spinor is pure 
\begin{equation}
  \label{eq:sigma-c-equations}
 \Sigma_{\BC} = \{ x \in \BR^{10} | u^M(x)=0  \}.
\end{equation}
If we find $\Sigma_{\BC} \in \BR^{10}$ then we get $\Sigma$ simply by intersecting $\Sigma_{\BC}$ with 
the space-time $\BR^{4}_\spt \subset \BR^{10}$. 

To solve the pure spinor equations (\ref{eq:pure-condition}) it is convenient to 
identify the $\Spin(10)$ spinor representation $S \simeq \BC^{32}$ with 
the space of anti-holomorphic $(0,p)$ forms, $p = 0, \dots, 5$, on the vector space $\BC^{5}\simeq \BR^{10}$.

The spinor $\ve(x)$ is pure at the origin. We use it to define a complex structure on 
the vector space $\BR^{10}$, so in the following we assume $\BR^{10} \simeq \BC^{5}$
where the isomorphism is defined by the pure spinor $\ve_{s}$. 

Given a pure spinor $\ve_{s}$, the spinor representation $S$ of $\Spin(10)$ can be constructed
as a Fock space using  action of the gamma-matrices.  As was explained around formula (\ref{eq:pure-definition}) we use the conventions such that the spinor $\ve_{s}$ is annihilated by the anti-holomorphic vectors $v^{\bar I}$. 
In the following, we use the indices $I, \bar I = 1\dots 5$ to denote the holomorphic 
and anti-holomorphic coordinates $x^{I} , x^{\bar I}$ on $\BC^{5} \simeq
\BR^{10}$. (Note that if $x^{I},x^{\bar I}$ are coordinates of a point in the original real space $\BR^{10} \simeq \BC^{5}$ 
then $x^{\bar I}$ is a complex conjugate of $x^{I}$. However, on 
the  complexified space $\BR^{10} \otimes \BC =\BC^{10}$  we use coordinates
 $x^{I}, x^{\bar I}$ as indendent.)
From our definition of the complex structure 
\begin{equation}
  \label{eq:annihilation}
  v^{\bar I} \gamma_{\bar I}\ \ve_{s} = 0 \quad \text{for any} \quad v \in V^{(0,1)}\
\end{equation}
we get that $\ve_{s}$ is annihilated by matrices $\gamma_{\bar I}, \bar I = 1 \dots 5$. 

Let us fix our notations more precisely. The $32\times 32$ Dirac  gamma-matrices representing  the
Clifford algebra on the space $\BR^{10}$ satisfy the canonical anticommutation relations
\begin{equation}
  \label{eq:anti-commutation}
 \{ \gamma_{M}, \gamma_{N} \} = 2 g_{MN}\ ,
\end{equation}
where $g_{MN}=\delta_{MN}$ is the standard unit metric on $\BR^{10}$ . 

Given the complex structure $J$ on $\BR^{10}$ compatible with the metric $g_{MN}$, we get a Hermitian metric $g_{I \bar J}$ 
on the complexified space $\BC^{10} = \BR^{10}\otimes \BC $ and then a
structure of the Clifford algebra on 
$\BC^{10}$. If $(x^{I}, x^{\bar I})$
are the coordinates on $\BC^{10}$, the corresponding basis
elements of Clifford algebra are represented by the matrices $\gamma_{I}, \gamma_{\bar I}$.
Moreover, since $g_{IJ} = g_{\bar I \bar J} = 0$ we have
\begin{equation}
  \label{eq:commutation-relations-gamma-holomorphic}
  \{ \gamma_{I}, \gamma_{\bar J}  \}  = 2 g_{I \bar J}\ , \quad 
  \{ \gamma_{I}, \gamma_{J} \}  = 0\ , \quad
  \{ \gamma_{\bar I}, \gamma_{\bar J} \}  = 0\ .
\end{equation}

We can use the inverse metric to raise indexes and then define gamma-matrices with the upper index
\begin{equation}
  \label{eq:gamma-upper-lower-relation}
  \gamma^{I} = g^{I \bar J} \gamma_{\bar J}\ , \quad \gamma^{\bar I} = g^{\bar I J} \gamma_{J}\ ,
\end{equation}
where $g^{I \bar K} g_{\bar K J} = \delta^{I}_{J}$.
Then 
\begin{equation}
  \label{eq:anti-commutation-relations-up-lower-index}
  \{ \gamma^{I}, \gamma_{J}  \}  = 2 \delta^{I}_{J}\ , \quad
  \{ \gamma^{\bar I}, \gamma_{\bar J} \}  = 2 \delta^{\bar I}_{\bar J}\ , \quad
  \{ \gamma^{\bar I}, \gamma_{J} \}  = \{ \gamma^{I}, \gamma_{\bar J} \} = 0.
\end{equation}

The construction of the spin representation $S$ as a Fock space is straightforward. 
\newcommand{\vac}{|\ve_s\rangle}
We define the \emph{vacuum state} $\vac$ as a state annihilated by 
all anti-holomorphic vectors in $V^{(0,1)}\subset \BC^{10}$ under the Clifford action (compare with (\ref{eq:annihilation}))
\begin{equation}
  \label{eq:definition-vacuum-state-pure-spinor}
  v^{\bar I} \gamma_{\bar I} \vac = 0 \quad \text{for all $(0,1)$ vectors $v$}.
\end{equation}

It  will be more convenient to use the $p$-forms instead of $p$-vectors in what follows and  
we use the Hermitian metric $g_{I \bar J}$ to identify $V^{(0,1)}$ with the space of holomorphic one-forms $V^*_{(1,0)}$.
Then 
\begin{equation}
  \label{eq:definition-vacuum-state-pure-spinor1}
  v_{I} \gamma^{I} \vac = 0 \quad \text{for all $(1,0)$ forms $v$}.
\end{equation}
We call $\gamma^{I}$ the lowering operators and $\gamma^{\bar I}$ the raising operators. 
The Fock space as a vector space is  spanned on the states (with $n=5$ in our case)  
\begin{equation}
  \label{eq:fock-space}
  \gamma^{\bar I_1 \cdots \bar I_k} \vac\ , \quad I_1 < I_2 < \cdots < I_k\ , \quad k \leq n.
\end{equation}
Let $\rho_p$ denote an antisymmetric $(0,p)$-form 
\begin{equation}
  \label{eq:rho-form-index-notation}
  \rho_p = \sum_{\bar I_1 < \bar I_2 < \dots < \bar I_p} \rho_{\bar I_1 \dots \bar I_p} \gamma^{\bar I_1 \dots \bar I_p}.
\end{equation}
Then an arbitrary spinor $\ve$ as a state in Fock space can be  written as 
\begin{equation}
  \label{eq:spinor-fock-space}
  \ve = \sum_{p=0}^{n} \rho_p  \vac.
\end{equation}
The space of anti-holomorphic forms $\oplus_{p} V^*_{(0,p)}$ is isomorphic to the spin representation space $S$. 
There is a natural $\BZ_{2}$ grading  on $S$ that
is compatible with the action of the generators $\gamma_{MN}$ of $\Spin(2n)$.
This $\BZ_2$ grading defines the chiral decomposition $S = S^{+} \oplus S^{-}$.
The space  $S^{+}$ of spinors of positive chirality
is the space of  forms of even degree $p$ 
and the space $S^{-}$ of spinors of negative chirality
is the space of  forms of odd degree $p$.

If $n$ is odd, then the representation $S^{+}$ and $S^{-}$ are dual to each other,
which means that there is a natural 
$\Spin(2n)$-invariant pairing between $S^{+}$ and $S^{-}$. 
If $\rho \in S^{+}$ and $\sigma \in S^{-}$, in the conventional
spin index notations the pairing is simply $\rho^\alpha \sigma_\alpha $. The same contraction in Fock space representation (\ref{eq:spinor-fock-space}) is
\begin{equation}
   \label{eq:pairing}
   (\rho, \sigma) : = (R[\rho] \wedge \sigma)_{top}.
\end{equation}
Here $|_{top}$ stands for picking up the coefficient of the top degree form normalized by some fixed element in $V^*_{(0,n)}$, 
and  $R[\rho]$ denotes the \emph{reverse order operation} on $S^{+}$, see e.g. \cite{MR1636473,MR0060497}
\begin{equation}
\begin{aligned}
  R[ \rho_p ] &= \rho_p \quad \text{for} \quad p = 4k, 4k+1,  \\
  R[ \rho_p ] &= -\rho_p \quad \text{for} \quad p = 4k+2, 4k+3.
\end{aligned}
\end{equation}

For $n=5$ the pairing between spinor $\rho = \rho_0 + \rho_2 + \rho_4 \in S^{+}$ and
spinor $\sigma = \sigma_1 + \sigma_3 + \sigma_5 \in S^{-}$
is 
\begin{equation}
  \label{eq:n=5=shevalle}
   (\rho,\sigma) = (\rho_0 \wedge \sigma_5 - \rho_2 \wedge \sigma_3 + \rho_4 \wedge \sigma_1).
\end{equation}

At the next step we rewrite the pure spinor condition for a spinor $\ve \in S^+$ 
\begin{equation}
  \label{eq:spinor-rho-notations}
  \ve = (\rho_0 + \rho_2 + \rho_4)\vac\ ,
\end{equation}
in terms of the constraints on the forms $\rho_0, \rho_2, \rho_4$.
In general, given a vector space $V = \BR^{2n}$,
and a complex structure on $V$, a pure spinor is a vacuum state
in the spin representation constructed as a Fock space. 
In other words, $\ve \in S$ is a pure spinor if it is annihilated by 
a half-dimensional isotropic subspace $L \subset V_{\BC}$ with $L \cap \bar L = 0$.\footnote{
Isotropic means that $g(L,L) = 0$ i.e. $g_{IJ}=g_{\bar I \bar J}=0$.}
A choice of $L\subset V_{\BC}$ defines a complex structure on $V$ by declaring $L$ to 
be the space of anti-holomorphic vectors $L = V^{(0,1)}$.  To summarize,
the space of complex structures on $V$ is isomorphic to the space
of equivalence classes of 
pure spinors $\ve$ modulo rescaling $\ve \sim \lambda \ve, \ve \in \BC^{*}$. 

%(see e.g.\cite{Berkovits:2001rb})
As we already mentioned above, if $n=5$ a spinor $\ve$ is a pure if and only if
\begin{equation}
  \label{eq:another-definition-n=5}
  \ve \Gamma_M \ve = 0, \quad M = 1,\dots,10.
\end{equation}
Now we rewrite (\ref{eq:another-definition-n=5}) using the form notation  (\ref{eq:spinor-rho-notations}) and (\ref{eq:pairing}) 
\begin{equation}
\label{eq:form-vector-condition-on-ve}
\begin{aligned}
 \ve v_{\bar I} \gamma^{\bar I} \ve  &=  0,\quad v \in V_{0,1}^{*} \\
 \ve v^{\bar I} \gamma_{\bar I} \ve &= 0,\quad v \in V^{0,1}.
\end{aligned}
\end{equation}

To simplify notations in the calculation we notice that for any spinor $\ve = \rho \vac$, where $\rho$ is a polyform, 
we have
\begin{equation}
  \label{eq:action-by-form}
  v_{\bar I} \gamma^{\bar I} \ve  = (v \wedge \rho) \vac\ ,
\end{equation}
where $v \wedge \rho$ denotes the usual external product of the antisymmetric forms $v$ and $\rho$.
Similarly, using (\ref{eq:gamma-upper-lower-relation}) we also have
\begin{equation}
  \label{eq:action-by-vector-field}
  v^{\bar I} \gamma_{\bar I} \ve = (2 i_v \rho) \vac\ ,
\end{equation}
where $i_v \rho$ denotes a contraction of the vector $v$ and a polyform $\rho$. 

We want to express the condition that a spinor is pure spinor as a constraint on $\rho$. 
After contracting (\ref{eq:action-by-form})  with $\langle\ve|$ we get (first equation of (\ref{eq:form-vector-condition-on-ve}))
\begin{equation}
  \rho_0 \wedge v \wedge \rho_4 - \rho_2 \wedge v \wedge \rho_2 + \rho_4 \wedge v \wedge \rho_0 = 0 \ ,
\end{equation}
for any 
anti-holomorphic one-form $v$, which means that if $\rho$ is pure then  
\begin{equation}
  \label{eq:rho0rho4-rho2-condition}
  \rho_0 \rho_4 = \frac 1 2 \rho_2 \wedge \rho_2.
\end{equation}
Similarly, the  second equation of (\ref{eq:form-vector-condition-on-ve}) 
implies that if $\rho$ is pure then  
\begin{equation}
  \label{eq:rho2rho4-condition}
  \rho_2 \wedge i_v \rho_4 = \rho_4 \wedge  i_v \rho_2 \quad \text{for any vector $v$}.
\end{equation}
Since  $0 = i_v(\rho_2 \wedge \rho_4) = i_v \rho_2 \wedge \rho_4 + \rho_4 \wedge i_v \rho_2$, we get that (\ref{eq:rho2rho4-condition}) is equivalent to 
\begin{equation}
  \label{eq:rho2rho4-condition-simpler}
  i_v \rho_2 \wedge \rho_4 = 0 \quad \text{for any vector $v$}.
\end{equation}
Notice that if $\rho_0 \neq 0$, the condition (\ref{eq:rho0rho4-rho2-condition})
implies (\ref{eq:rho2rho4-condition}). Indeed, it is easy to check that $\rho_2 \wedge \rho_2 \wedge i_v \rho_2$ vanishes
identically in five dimensions for any two-form $\rho_2$ and vector $v$.

Another way to derive the pure spinor constraint (\ref{eq:rho0rho4-rho2-condition}) is 
to notice  that all pure spinors $\rho$ with $\rho_0 \neq 0$,
modulo rescalings $\rho \to \lambda \rho, \lambda \in \BC^{*}$
are in the $\Spin(10)$ orbit of the vacuum spinor  $\vac$.
The $\Spin(10)$ acts on $S^{+}$ as
\begin{equation}
  \vac \mapsto \exp ( \omega_{\bar I\bar J} \gamma^{\bar I\bar J}) \vac\ .
\end{equation}
(We write only $(0,2)$ components  $\omega_{\bar I\bar J}$ of all $\Spin(10)$ generators,
because $\vac$ is annihilated by holomorphic generators $\gamma^{I}$). 
Then 
\begin{equation}
  \label{eq:spin5c-rotation-of-vacuum}
  \vac \mapsto (1 + \omega_{\bar I\bar J} \gamma^{\bar I\bar J}  +  \frac 1 2 \omega_{\bar I\bar J} \omega_{\bar K\bar L}
 \gamma_{\bar I \bar J}\gamma_{\bar K\bar L})\vac,
\end{equation}
which can be rewritten as 
\begin{equation}
 \vac \mapsto (1 + \omega_2 + \frac 1 2 \omega_{2} \wedge \omega_{2}) \vac.
\end{equation}
Here $\omega_2$ is a two-form $\omega_2 = \omega_{\bar I\bar J} \gamma^{\bar I\bar J}$. 
Hence, all pure spinors with $\rho_0 \neq 0$  can be
parametrized by a scale factor $\tilde{\rho}_0 \in \BC$ and
a two-form $\omega_2 \in \Lambda^2 (\BC^5)$. (This is a well-known local parametrization
of pure spinors in ten dimensions used in \cite{Berkovits:2000fe,Berkovits:2002zk}).
In the $\rho=\rho_0 + \rho_2 + \rho_4$ is expressed in terms 
of $\tilde{\rho}_0$ and $\omega$ as 
\begin{equation}
  \rho_0 = \tilde{\rho}_0\ , \quad 
  \rho_2 = \tilde{\rho}_0 \omega_2\ , \quad
  \rho_4 = \frac 1 2 \tilde{\rho}_0 \omega_2 \wedge \omega_2.
\end{equation}
The quadratic constraints (\ref{eq:rho0rho4-rho2-condition}) are satisfied.

\subsection{\label{se:pure-spinor-surface}Pure spinor surface in $\BR^{10}$}
Now we are ready to rewrite the conformal Killing spinor (\ref{eq:conformal-Killing-new-system}) in the form notations on $\BC^{5}$
and solve the pure spinor constraint (\ref{eq:rho0rho4-rho2-condition}) and (\ref{eq:rho2rho4-condition-simpler}). 

We use the Fock space representation of $S^-$ to identify the superconformal generator $\ve_{c}$ 
with  three anti-holomorphic forms $\vv,\mm,\ww$, where $\vv$ is a $(0,1)$-form, $\mm$ is a $(0,3)$ and $\ww$ is a $(0,5)$-form on $\BC^5$
(clearly, the total number of components matches as $5+10+1 = 16$). More explicitly
\begin{equation}
  \label{eq:form-notations}
  \ve_{c}  = \left(\vv_{\bar I} \gamma^{\bar I} + \frac 1 {3!} \mm_{\bar I_1 \bar I_2 \bar I_3} \gamma^{\bar I_1 \bar I_2 \bar I_3}
+ \frac {1} {5!} \ww_{\bar I_1 \bar I_2 \bar I_3 \bar I_4 \bar I_5} \gamma^{\bar I_1 \bar I_2 \bar I_3 \bar I_4 \bar I_5}\right)\ve_{s}\ .
\end{equation}

A conformal Killing spinor (\ref{eq:conformal-killing-solutions}) formally extended to $\BR^{10} = \BC^5$ is then 
\begin{equation}
\label{eq:killing-form-notation}
\begin{split}
  \ve(x) = \ve_{s} + (\xi_{\bar J} \gamma^{\bar J} + x^{\bar I} \gamma_{\bar I}) \ve_{c} \\
= (( 1 +  2 i_x  \vv)  + (\xi \wedge \vv + 2 i_x \mm) + (\xi \wedge \mm + 2 i_x \ww)) \vac,
\end{split}
\end{equation}
where we introduced the $(0,1)$ one-form $\xi_{\bar I} = g_{\bar I J} x^{J}$. 

If $x \in \BR^{10}$, so the coordinates $x^{M}$ are real, 
then $x^{I}$ and $x^{\bar I}$ are complex conjugate to each other. 
In this case the $(0,1)$ form $\xi_{\bar I}$ and the $(0,1)$ vector $x^{\bar I}$ are related through complex conjugation. More generally, one can treat $\xi_{\bar I}$
and $x^{\bar I}$ as independent, which corresponds to taking complex $x^{M}$.

Recall that we defined $\Sigma_{\BC} \subset \BR^{10}$ as a set of points where the spinor
$\ve(x)$  (\ref{eq:killing-form-notation}) is pure. Clearly, the point $x^{M} = 0$ is always in $\Sigma_{\BC}$. We say that $\Sigma_{\BC}$ is non-trivial if $x^{M}=0$ 
belongs to a component of positive dimension.

We call a $(0,3)$ form $\mm$ totally decomposable 
if there exist three $(0,1)$-forms $\mu_1,\mu_2,\mu_3$ such
that $\mm = \mu_1 \wedge \mu_2 \wedge \mu_3$.

Now we formulate the key result of this section.

{\bf Proposition.} Given a pure spinor $\ve_{s}$, a  pure spinor
 hypersurface $\Sigma_{\BC} \subset \BR^{10}$ is non-trivial if
and only if 
$\ve_{c}$ in parametrization of (\ref{eq:form-notations}) satisfies  $\ww = 0$
and $\mm$ is totally decomposable.
In this case the hypersurface $\Sigma_{\BC}$ is described by the equation
\begin{equation}
  \label{eq:equation-sigma}
  ( \xi + 2(\xi, x) \vv ) \wedge \mm = 0\ ,
\end{equation}
where the complex coordinates $(x^{\bar I}, \xi_{\bar I} = g_{\bar I J} x^{J})$ 
are defined by the complex structure on $\BR^{10}$ 
associated to the pure spinor $\ve_{s}$.

We delegate the proof that the non-trivial $\Sigma_{\BC}$ requires 
$\ww=0$ and $\mm$ to be decomposable to the Appendix \ref{se:app-decomp}.
Here  we just show that if both conditions are satisfied $\Sigma_{\BC}$ is given by (\ref{eq:equation-sigma}).

For the spinor (\ref{eq:form-notations}) the quadratic pure spinor constraint
(\ref{eq:rho2rho4-condition-simpler}) with $v = x$ takes the form 
\begin{equation}
\label{eq:main-mu-xi-v=0}
0 = i_x \rho_2 \wedge \rho_4 =  i_x ( \xi \wedge \vv) \wedge (\xi \wedge \mm) = (x,\xi) \vv \wedge \xi \wedge \mm\ .
\end{equation}
For a real non-zero $x$ the pairing $(x,\xi)=\frac 1 2 |x|^2$ is also non-zero. 
Therefore $\vv \wedge \xi \wedge \mm$ must vanish and consequently 
\bea
\label{eq:zero-expand}
0=i_x(\xi\wedge \vv\wedge \mm)=i_x \xi \ \vv \wedge \mm-i_x \vv \ \xi\wedge \mm+ \xi\wedge \vv \wedge i_x\mm\ .
\eea

Now we proceed with the constraint
 (\ref{eq:rho0rho4-rho2-condition})
\begin{equation}
  \label{eq:rho4rho0-rho2-with-w=0}
  (1 + 2 i_x \vv) \wedge (\xi \wedge \mm) = \frac 1 2 (\xi \wedge \vv + 2 i_x \mm)^2\ .
\end{equation}
First we expand both sides  
\begin{equation}
  \label{eq:lem2-exp1}
 \xi \wedge \mm + 2 i_x \vv \wedge \xi \wedge \mm =  \frac 1 2 (\xi \wedge \vv)^2 + 2  \xi \wedge \vv  \wedge i_x \mm + 
2(i_x \mm)^2\ ,
\end{equation}
and notice that $(\xi \wedge \vv)^2 = 0$ and also $(i_x \mm)^2 = 0$ because we assume that $\mm$ is totally decomposable.

Together with (\ref{eq:zero-expand}) the equation  (\ref{eq:lem2-exp1}) reduces to
\begin{equation}
  \label{eq:main-xi-mu-2}
  (\xi + 2 (x,\xi) \vv ) \wedge \mm = 0\ .
\end{equation}
Since (\ref{eq:main-xi-mu-2}) imply $\xi\wedge \vv\wedge \mm=0$ we conclude that
if $\ww = 0$ and $\mm$ is totally decomposable the pure spinor constraints (\ref{eq:rho0rho4-rho2-condition}),(\ref{eq:rho2rho4-condition-simpler}) are equivalent to (\ref{eq:main-xi-mu-2}).

Now let us solve the equation (\ref{eq:main-xi-mu-2}) for $\Sigma_{\BC}$. 
There are only two topologically distinct cases: $\mm = 0 $ and $\mm \neq 0$. 
If $\mm = 0$ the equation (\ref{eq:equation-sigma}) for $\Sigma_{\BC}$ is trivial and $\Sigma_{\BC} = \BR^{10}$, $\Sigma=\BR^4\_{\spt}$.
If $\mm = \mu_1 \wedge \mu_2 \wedge \mu_3 \neq 0$ then it is convenient to choose an 
orthonormal coordinate system $z_1,..,z_5$ in $\BC^5 \cong \BR^{10}$ such
that 
\bea
\label{eq:mu-3form}
\mm = \mu\ \overline {dz_1} \wedge \overline {dz_2} \wedge \overline {dz_3}
\eea
with $\mu \in \BC^{*}$.

Orthonormality of the chosen coordinate system implies that 
$g_{I \bar J} = g_{\bar I J} = \frac 1 2$. In this coordinates the equation for $\xi_{\bar I}$ is  
\begin{equation}
\label{eq:xiv-s6-in-basis}
\begin{aligned}
\xi_{\bar 4} = - |\xi|^2 \vv_{\bar 4}\ , \\
\xi_{\bar 5} = - |\xi|^2 \vv_{\bar 5}\ ,
\end{aligned}
\end{equation}
where $|\xi|^2 = 2 g^{I \bar J} \overline{ \xi_{\bar I}} \xi_{\bar J}=x^2$. 

If $\vv_{\bar 4} = \vv_{\bar 5} = 0$ then $\Sigma_{\BC}$ is a complex three-plane 
$\Sigma_{\BC} = \BC^{3}$ defined by $z^4 = z^5  =0$.  
Otherwise, $\Sigma_{\BC}$ is  a real six-dimensional sphere $\Sigma_{\BC} = \Sph^{6}$
defined by the equations (\ref{eq:xiv-s6-in-basis}). 
For  illustration, consider an example when 
$z_1,\dots, z_5$ are related to the original coordinates $x_1,\dots,x_{10}$ on $\BR^{10}$ in the simplest way\footnote{Notice that we have chosen the simplest relation just to illustrate the idea. In general the relation between the original basis $x^M$ and the complex basis $z^I$
that diagonalize $\mm$ could be different.}
\begin{equation}
z^I=x^{2I-1}-i x^{I}.
\end{equation}
Then the equations (\ref{eq:xiv-s6-in-basis}) can be written in real notations as follows
\begin{equation}
  x^a + x^2 \vv^a = 0\ ,\quad a=7\dots 10,
\end{equation}
and  $x^2=x^M x_M$. The sphere $\Sph^6$ is located inside the  $\BR^{7}$
spanned by first six directions in $\BR^{10}$ and the vector $\vv^a$.

Now, depending on the relative
orientation of the space-time $\BR^{4}_\spt \subset \BR^{10}$
and $\Sigma_{\BC} \subset \BR^{10}$, we obtain
various pure spinor hypersurfaces $\Sigma =\Sigma_{\BC}\bigcap \BR^{4}_\spt$.
In the example above $\Sigma$ is just a point $x^M=0$ i.e. it is trivial, but in general $\Sigma=\Sph^n$
with $n=1,\dots,3$ or $\Sigma=\BR^n$ with $n=1,\dots,4$.

Let us  summarize possible cases for $\Sigma$:
\begin{enumerate}
\item {If  $\mm=0$ then $\Sigma_{\BC} = \BR^{10}$. Then automatically  $\Sigma = \BR^4_\spt$. }

\item {If  $\mm \neq 0 $  but $\vv \wedge \mm  = 0$ then
 $\Sigma_{\BC} = \BR^{6}$. Then $\Sigma$ 
is $\BR^n$, where $n=1,\dots,4$ depending on the relative orientation of
$\BR^{4}_{\spt}$  and $\Sigma_{\BC}$.}
\item {If $\mm \neq 0$ and $\vv \wedge \mm  \neq 0$ then 
$\Sigma_{\BC} = \Sph^{6}$. Then $\Sigma$ 
is $\Sph^n$, where $n=1,\dots,3$ depending on the relative orientation
of $\BR^{4}_{\spt}$ and $\Sigma_{\BC}$.}
\end{enumerate}

The third case could be related to the second one by a suitable
conformal transformation as explained in section \ref{se:mneq0}.

\subsection{\label{se:almost-complex-structure-on-pure-spinor-hypersurface} Complex structure on the pure spinor hypersurface}

We have just shown that for a suitable choice of 
spinors $(\ve_{s}, \ve_{c})$ the  supersymmetry spinor $\ve(x)$ is pure on a
 hypersurface $\Sigma_{\BC} \subset \BR^{10}$. If $\Sigma_{\BC}$ is non-trivial 
then $\Sigma_{\BC}$ is either $\BR^{10}$, $\BR^{6}$ or $\Sph^{6}$.

In the previous section we used the pure spinor $\ve_s=\ve|_{x=0}$ to define complex structure on $\BR^{10}$ as on the vector space (not as on the manifold $\BR^{10}$). 
 It was merely a technical trick that helped us to find $\Sigma_{\BC}$. 
Now, when this is done, we will find an almost complex structure 
$J(x) \in \End(\BR^{10},\BR^{10})$  at each point $x$ on $\Sigma_{\BC}$
defined by $\ve(x)$.  
The complex structure at the origin $J(x=0)$ coincides with the 
base complex structure on $\BR^{10}$ defined by $\ve_{s}$ and
used in the previous section.

This complex structure $J(x)$ or, more precisely, the space of anti-holomorphic vectors
$V^{(0,1)}_x$ at each point $x$ defines locally the space of allowed
 supersymmetric combinations of the contour directions $v^\mu$ and the scalar couplings $v^A$ of the Wilson loop (\ref{eq:extended-Wilson-loop-definition}).

Let $Z^{M}_{\bar I}$ where $M = 1 \dots 10$,  $I,\bar I = 1 \dots 5$ 
be $x$-dependent $10 \times 5$ basis matrix of $V_x^{(0,1)}$.
Similarly, let $Z^{M}_{I}$ be the basis matrix of $V_x^{(1,0)}$,
so 
\begin{equation}
\label{eq:Zx}
x^M=Z^M_I x^I+Z^M_{\bar I} x^{\bar I}.
\end{equation}

The matrix $Z^{M}_{\bar I}$  defines the anti-holomorphic 
vector space $V_x^{(0,1)}$ associated with the pure spinor $\ve(x)$ at a
 given point $x\in\Sigma_{\BC}$
\begin{equation}
  \label{eq:Z-annihilate-epsilon}
   Z^{M}_{\bar I}(x) \gamma_{M} \ve(x) = 0.
\end{equation}
We can normalize $Z_I^M$ as
\begin{equation}
\label{eq:Znorm}
\delta_{MN}Z^M_I Z^N_J=0\ ,\quad \delta_{MN}Z^M_I Z^N_{\bar J}=g_{I\bar J}.
\end{equation}

We assume for now that $\vv \wedge \mm = 0$ which means that $\Sigma_{\BC}$ is either 
the total space $\BR^{10}$ if $\mm$ vanishes, or a six-plane $\BR^{6} \subset \BR^{10}$ if $\mm$ is a non-zero decomposable three-form.
The supersymmetry spinor $\ve(x)$  (\ref{eq:killing-form-notation}) is explicitly  given by 
\begin{equation}
  \label{eq:killing-explicit-for-muv=0}
  \ve(x) = (1 + 2 x^{\bar I} \vv_{\bar I}) \left(1 + \frac  1 2 \alpha_{\bar I \bar J} \gamma^{\bar I \bar J}\right) \ve_{s}
\end{equation}
where $(1+2 x^{\bar I} \vv_{\bar I})$ is a scalar multiplier and the $(0,2)$ form $\alpha_{\bar I \bar J}$ is 
\begin{equation}
  \label{eq:alpha}
  \alpha_{\bar I \bar J}(x)  = \frac {\xi_{\bar I} v_{\bar J} - \xi_{\bar J} v_{\bar I} + 2 x^{\bar K} \mm_{\bar K \bar I \bar J}} { 1+2 x^{\bar I} \vv_{\bar I} }.
\end{equation}
To find  $Z^{M}_{\bar I}(x)$ we start with (\ref{eq:Z-annihilate-epsilon}) at $x=0$
  
\begin{equation}
  %\label{eq:Z-annihilate-epsilon}
  \hat Z^{M}_{\bar I} \gamma_{M} \ve_{s} \equiv \gamma_{\bar I}\ve_s = 0,
\end{equation}
where ${\hat Z}^M_{\bar I}=Z^M_{\bar I}(x=0)$, 
and multiply it by $(1 + \frac  1 2 \alpha_{\bar I \bar J} \gamma^{\bar I \bar J})$ from the left. Then we use
the anticommutation relations to move this factor to the right 
and also the fact that $\alpha \wedge \alpha = 0$ to express $\ve_s$ through  $\ve(x)$. As a result we get 
\begin{equation}
\label{eq:ZZ}
  Z^{M}_{\bar I}(x) = \hat Z^{M}_{\bar I} + 2  \hat Z^{M}_{K}g^{K \bar J}\alpha_{\bar J \bar I}(x).
\end{equation}

\section{Classification of the $\SO(5,1)\times \SO(6)$ orbits in the space of superconformal charges
\label{se:classification-loops}}
In section \ref{se:pure-spinor-manifold-sigma} we found the conditions on a pair of spinors $(\ve_s,\ve_c)$  such that the
 conformal Killing spinor $\ve(x)$ (\ref{eq:conformal-Killing-spinor}) is pure 
on a nontrivial hypersurface $\Sigma\in \BR^4$. 
The Wilson loop operator (\ref{eq:extended-Wilson-loop-definition}) on $\Sigma$
 is supersymmetric with respect to $\ve(x)$
if $v^M$ is anti-holomorphic at each point $x\in\Sigma$ with respect to  the complex
structure $J(x)$. That means $v^M(X)= Z_{\bar I}^M(x) v^{\bar I} $ for
a suitable $v^{\bar I}$.  
To find a real contour in $\BR^{4}_{\spt}$  one is compelled to choose $v^{\bar I}$
such that $v^\mu$ is real and $v^\mu=dx^\mu/ds$ for some contour
$\gamma:x^\mu(s)\subset \Sigma$. Using the matrix
$Z^M_{\bar I}$, introduced in the previous section, one can construct all possible
supersymmetric Wilson loop operators on $\Sigma$.

 It is clear nevertheless that this description 
is not unique in a sense that different operators can be related to each
other by the action 
of the global symmetry
group. For
example if we start with some pair $(\ve_s,\ve_c)$,  
that leads to a nontrivial $\Sigma$, we can always move the origin of the
coordinate system and obtain a new pair $(\ve_s',\ve_c')$. Therefore the same
contour on $\Sigma$ and hence the corresponding  Wilson loop operator will be
described twice, once as corresponding to  $(\ve_s,\ve_c)$ and another time as
corresponding to $(\ve_s',\ve_c')$. To avoid double-counting we
should factorize the space of suppersymmetric Wilson operators  by the shifts in
$\BR^4_\spt$, and in general by the total  global bosonic symmetry group
of the theory  $SO(5,1)\times SO(6)$, where $\SO(5,1)$ is the conformal 
group of one-point compactification of $\BR^{4}_{\spt}$ and $\SO(6)$ is
the R-symmetry group. 

Let us notice that partially we have already fixed the ``conformal gauge'' by requiring
that $\ve_s$ is pure i.e. the  origin of coordinate system
belongs to $\Sigma$.  
Clearly this is not enough as other symmetries including shifts along $\Sigma$
and conformal transformations still have to be gauged away.     
Ultimately, we would like 
to find the space of all equivalence classes of pairs $(Q_{\ve},W)$
modulo the global symmetry. 

In this paper we consider only an interesting subclass of the pairs
$(Q_{\ve},W)$
the pure-spinor case of this problem, 
i.e. the pairs when the contour of $W$ is located on a pure-spinor surface $\Sigma$
and the couplings on $W$ are defined by anti-holomorphic vectors. 

The problem of finding equivalence classes in the other, 
not pure-spinor case, when contour of $W$ is just an orbit
of conformal transformation generated by $Q$, is left for the future. 
As we have mentioned in the introduction, if we require that the orbits
are compact, there are no other curves except simple generalization 
of circle known as $\frac p q$ Lissajous figure. For $x^{\mu}$ being coordinates
on $\BR^{4}$, take the orbit $x_1 + ix_2 = r_1 e^{i p t}, x_3+ i x_4
= r_2 e^{i q t}$ corresponding to generator of the $\SO(2) \oplus \SO(2)$
rotations of the 12-plane and the 34-plane, such that  $\frac p q \in \BQ$.

%  supercharges and all Wilson loops. Indeed the supercharges $(\ve_s,\ve_c)$ that give rise to the pure spinor $\ve(x)$ is a special sub-case of a general case. In this paper we focus on a more interesting task of factorizing the pairs $(\ve_s,\ve_c)$ leading to pure spinors while leaving the task of factorizing the whole space of supercharges and the corresponding Wilson lines (\ref{eq:generic-point},\ref{eq:gener-Wilson-local}) for the future. 

 The bosonic global symmetry group of the $\CalN=4$ SYM on $\BR^{4}_{\spt}$
 is the product of
 the conformal group $\SO(5,1)$ of the four-dimensional Euclidean space $\SO(5,1)$ and 
 the R-symmetry group $\SO(6)$. 

Actually, to classify pairs $(Q,W)$ in a meaningful way we
should say more precisely that $Q$ denotes one-dimensional fermionic subspace
of the superconformal algebra. In other words, if $Q_{\ve}$ is a symmetry 
of $W$ then so obivously is a $Q_{\lambda \ve}, \lambda \in \BC^{*}$. 
When we represent $Q$ by a pair of spinors $(\ve_{s}, \ve_{c})$
we actually consider equivalence classes 
under the action of $\SO(5,1) \times \SO(6)\times \BC^*$ 
on this space, where $\BC^*$ acts by a simple rescaling $(\ve_s, \ve_c) 
\to (\lambda \ve_s, \lambda \ve_c)$, $\lambda \in \BC^{*}$.

It is convenient to represent the action of $\SO(5,1) \times \SO(6)$ on the
space of pairs $(\ve_s,\ve_c)$ using the spinor representation of the $\SO(11,1)$ group acting on the $64$ component spinor 
that is built of $(\ve_s$, $\ve_c)$.  

Before we proceed with further details  let us explain how the conformal group
$\SO(5,1)$ acts on the $(\ve_s$, $\ve_c)$. First we compactify $\BR^4_\spt$ into $S^4$. The group $\SO(5,1)$
acts on $S^4$ as follows. Let $(1,2,3,4,11,12)$ be the set of indexes in the 
space $\BR^{5,1}$ where acts $\SO(5,1)$ canonically. Let us consider the $
SO(5,1)$-invariant cone 
\begin{equation}
  \label{eq:cone-in-six-dimensions}
  X_1^2 + X_2^2 + X_3^2 + X_4^2 + X_{11}^2 - X_{12}^2 = 0.
\end{equation}
%Obviously it is preserved by the action of $\SO(5,1)$. 
For any point on this cone and $X_{12} \neq 0$ 
the five dimensional vector $n_i = X_i / X_{12}$ has unit norm and therefore
parametrizes unit $S^4$ within $\BR^5$. The action of the conformal group $\SO(5,1)$ on $S^4$ is the action on $\vec{n}$ induced from the canonical action of $\SO(5,1)$ on $(X_1,\dots,X_{11},X_{12})$.

For example the generators $K_\mu$ of the special conformal transformation on $\BR^4$ are related to the generators of $\SO(5,1)$ as follows
\begin{equation}
  \label{eq:K-relation-SO5-1}
  K_\mu = - R_{11, \mu} + R_{12, \mu}\ .
\end{equation}

To check this we perform a special conformal transformation (\ref{eq:K-relation-SO5-1}) parametrized by vector $b_\mu=\frac {v_\mu} 2$. Without loss of generality we can choose $v_\mu$ to be along the direction $X_1$. Then the action of $\SO(5,1)$ is restricted on the directions $\{X_1,X_{11},X_{12}\}$. The corresponding generator 
\begin{equation}
  \label{eq:K-1-generator-explicit-matrix}
  K=
  \begin{pmatrix}
    0 & 1 & 0 \\
    1 & 0 &-1 \\
    0 & 1 & 0
  \end{pmatrix}
\end{equation}
can be exponentiated  as follows
\begin{equation}
  \label{eq:finite-K-1-special-conformal-transformation}
  e^{b K} =
  \begin{pmatrix}
    1 + \frac {b^2} 2 & b & -\frac {b^2}{2} \\
    b_1 & 1 & -b_1 \\
    \frac{b^2}{2} & b & 1 - \frac {b^2}{2}
  \end{pmatrix}.
\end{equation}
This matrix generates the transformation
\begin{equation}
  \label{eq:transformation-explicit}
  \begin{aligned}
  n_1 &\to \frac {n_1 + b(1-n_5)} { (1 + \frac {b^2} 2) + b n_1 - \frac{b^2}{2}n_5}\ ,  \\
  n_\mu &\to \frac {n_\mu} { (1 + \frac {b^2} 2) + b n_1 - \frac{b^2}{2}n_5}\ ,\quad \mu\neq 1\ ,  \\
  n_5 &\to \frac { (1-\frac{b^2} 2) n_5 + b n_1 + \frac{b^2}{2}}  { (1 + \frac {b^2} 2) + b n_1 - \frac{b^2}{2}n_5}\ .
  \end{aligned}
\end{equation}

Using the relation between the unit vector $n_\mu,n_5$ on $S^4 \subset \BR^5$
and the stereographic projective coordinates $x_\mu$ on $\BR^{4}$ 
%(here $r$ is the location of the equator)
%\bea
%  \label{eq:projective-coordinates-unit-vector}
%  x_\mu = n_\mu \frac {2}{1 + \frac{n_5}{r}}\ , \\
%  n_\mu = \frac{x_\mu} {1 + \frac {x^2}{4 r^2}}\ ,\quad 
%  n_5 = r \frac { 1 - \frac {x^2}{4r^2}} { 1 + \frac{ x^2}{4r^2}}\ , 
%\eea
% WE SET r=1
\begin{equation}
\begin{aligned}
  x_\mu &= n_\mu \frac {2}{1 + n_5}\ , \\
  n_\mu &= \frac{x_\mu} {1 + \frac {x^2}{4 }}\ ,\quad 
  n_5 &= r \frac { 1 - \frac {x^2}{4}} { 1 + \frac{ x^2}{4}}\ , 
\end{aligned}
\end{equation}
we get the usual formula for the special conformal transformations 
\begin{equation}
  \label{eq:special-conformal-arbitrary}
  x_\mu \to \frac { x_\mu + v_\mu x^2} { 1 + 2 v_\mu x_\mu + v^2 x^2}\ .
\end{equation}

At the next step we want to find the action of the conformal group $SO(5,1)$ on the conformal Killing spinor on $\BR^4_\spt$ (\ref{eq:conformal-killing-solutions}). It is defined as follows. If $u^\mu$ is a vector field generating a conformal transformations, then $\ve(x)$ transforms as
\begin{equation}
  \label{eq:transformation-of-epsilon-under-conformal-vector-field}
  \delta \ve = L_{u} \ve - \frac 1 2 \lambda \ve
\end{equation}
where $L_{u} \ve = u^{\mu} \partial_{\mu} \ve + \frac 1 4 \partial_{\mu} u_{\nu} \Gamma^{\mu \nu} \ve $ is the Lie derivative acting
on $\ve$ and $\lambda = \frac 1 4 \partial_{\mu} u^{\mu}$ is the conformal scaling factor. This formula follows from the fact, that the conformal Killing spinor
 $\ve$ under  conformal rescaling of metric  $g_{\mu\nu}\rightarrow e^{2 \Omega} g_{\mu \nu}$ 
transforms as $\ve\rightarrow e^{\Omega/2} \ve$.
To find the the action of the conformal group $SO(5,1)$ on the pair $(\ve_s,\ve_c)$ 
one may find vector field $u^\mu$ that corresponds to a generator $R_{mn}\in so(5,1)$, and then  find $(\delta \ve_s,\delta \ve_c)$ through (\ref{eq:transformation-of-epsilon-under-conformal-vector-field}).

As an example, consider the
 case of the special conformal transformation $-R_{\mu,11}+R_{\mu,12}$.
For an infinitesimal $v^\mu$ the corresponding vector field is $u^\mu=v^\mu x^2-2x^\mu (xv)$ as follows from (\ref{eq:special-conformal-arbitrary}). Then (\ref{eq:transformation-of-epsilon-under-conformal-vector-field}) implies that $\delta \ve_s=0$ and $\delta \ve_c=v^\mu \Gamma_\mu \ve_s$. This infinitesimal transformation can be easily integrated for a finite $v^\mu$
\begin{equation}
\label{eq:epsilon-special-conformal}
\begin{pmatrix}
\ve_s\\
\ve_c
\end{pmatrix} \rightarrow
\begin{pmatrix}
  1 & 0 \\
 \Gamma_{\mu} v^{\mu} & 1
\end{pmatrix}
\begin{pmatrix}
  \ve_{s} \\
  \ve_{c}
\end{pmatrix}.
\end{equation}
In the case of translation of $x^{\mu}$ by $v^{
\mu}$ one obviously gets 
\begin{equation}
\label{eq:epsilon-translation}
\begin{pmatrix}
\ve_s\\
\ve_c
\end{pmatrix} \rightarrow
\begin{pmatrix}
  1 & \Gamma_{\mu} v^{\mu} \\
  0 & 1
\end{pmatrix}
\begin{pmatrix}
  \ve_{s} \\
  \ve_{c}
\end{pmatrix},
\end{equation}
%This is in agreement with (\ref{eq:special-conformal-spinors}) since 
%\bea
%-\Sigma_{\mu,11}+\Sigma_{\mu,12}=\left(\begin{array}{cc}
%0 & 0 \\
%\Gamma_\mu & 0
%\end{array}\right)\ .
%\eea
The dilatations by factor $e^{\Omega}$ are represented as 
\begin{equation}
\label{eq:epsilon-dilatation}
\begin{pmatrix}
\ve_s\\
\ve_c
\end{pmatrix} \rightarrow
\begin{pmatrix}
  e^{\Omega/2} & 0 \\
  0 & e^{-\Omega/2}
\end{pmatrix}
\begin{pmatrix}
  \ve_{s} \\
  \ve_{c}
\end{pmatrix},
\end{equation}
and, finally, the space-time $\SO(4)$ rotations and the $\SO(6)$   R-symmetry transformations  are represented as
\begin{equation}
\label{eq:epsilon-rotation}
\begin{pmatrix}
\ve_s\\
\ve_c
\end{pmatrix} \rightarrow
\exp \begin{pmatrix}
  \frac 1 4 R_{MN} \Gamma^*_{[M} \Gamma_{N]} & 0 \\
  0 &   \frac 1 4 R_{MN} \Gamma_{[M} \Gamma^*_{N]}
\end{pmatrix}
\begin{pmatrix}
  \ve_{s} \\
  \ve_{c}
\end{pmatrix}.
\end{equation}

The spin representation (\ref{eq:epsilon-special-conformal})-(\ref{eq:epsilon-rotation}) of $\SO(5,1) \times \SO(6)$ can 
 can be embedded into the Clifford algebra of $\BR^{11,1}$ represented
 by the following $64\times 64$ gamma-matrices 
\begin{equation}
\begin{aligned}
  \label{eq:so11-1}
  \hat \gamma_{M} &=
  \begin{pmatrix}
    \gamma_M & 0 \\
    0 & -\gamma_M   
  \end{pmatrix}\ ,  \quad \quad M=1\dots 10\ , \\
\hat \gamma_{11} &=   \begin{pmatrix}
    0 & 1_{32\times 32} \\
    1_{32\times 32} & 0 
  \end{pmatrix}\ ,  \quad 
\hat \gamma_{12} =   \begin{pmatrix}
    0 & -1_{32\times 32} \\
    1_{32\times 32} & 0 
  \end{pmatrix}\ . 
\end{aligned}
\end{equation}
Then the $\SO(11,1)$ chirality operator is
\begin{equation}
  \label{eq:so11-1chirality}
\hat \gamma_{13} = -i \hat \gamma_1
\hat \gamma_2 \dots \hat \gamma_{12} =
\begin{pmatrix}
  1_{16 \times 16} & 0  & 0 & 0 \\
  0 & -1_{16 \times 16} & 0 & 0 \\
  0 & 0  & -1_{16 \times 16} & 0 \\
  0 & 0  &  0 & 1_{16 \times 16} 
\end{pmatrix}\ .
\end{equation}
Therefore the spinor 
\begin{equation}
  % \label{eq:so111positive-chirality}
   \label{eq:ve}
  \ve =
  \begin{pmatrix}
    \ve_s \\
    0 \\
    0 \\
    \ve_c 
  \end{pmatrix}\ ,
\end{equation}
is a $\SO(11,1)$ Weyl spinor of positive chirality, while the  $\ve_s$  
 and $\ve_c$ from (\ref{eq:ve}) are the $\SO(10)$ chiral Weyl spinors
of opposite chiralities. 
One can check that the action by the  conformal $\SO(5,1)$ group
and the $\SO(6)$ group on the conformal Killing spinor $\ve(x) = \ve_{s}
+ \gamma_{\mu} x^{\mu} \ve_{c}$
is represented precisely in the same way as $\SO(5,1) \times \SO(6)
\subset \SO(11,1)$ action on (\ref{eq:ve}).

We denote positive and negative chiral representations of $\SO(11,1)$ as
$S^{+}_{11,1}$ and $S^{-}_{11,1}$.
Now recall that the Weyl  representations of $\SO(10)$, which we called $S^{+}$ and $S^{-}$, are
related by complex conjugation. Namely, if
$\ve_{s}^{*}, \ve_{c}^{*}$ denote complex conjugates to
$\ve_{s},\ve_{c}$, then $\ve_{s}^*$ transforms in $S^{-}$ and $\ve_{c}^{*}$ transforms
in $S^{+}$. Another important observation is that the pair $(\ve_{s}^*,
\ve_{c}^*)$ transforms under $\SO(5,1) \times \SO(6)$ in the same way as the $\SO(11,1)$ Weyl spinor of negative chirality
\begin{equation}
  \label{eq:so11-1-negative}
  \tilde \ve  =
  \begin{pmatrix}
    0 \\
    \ve_{s}^{*} \\
    \ve_{c}^{*} \\
    0
  \end{pmatrix}\ .
\end{equation}

To classify the supercharges modulo $\SO(5,1) \times \SO(6)$ we
construct the $\SO(5,1) \times \SO(6)$ invariants in the spin space
$S^{+}_{11,1} \oplus S^{-}_{11,1}$ by contracting a twelve-dimensional spinor with a Diract conjugated one.
Thus for any pair of spinors $(\ve_1, \ve_2)\in S^{+}_{11,1} \oplus S^{-}_{11,1}$ we can construct a bilinear 
\begin{equation}
  \label{eq:bilinears}
  \rho_{i_1 \dots i_p j_1 \dots j_q} = \bar \ve_{1} \hat \gamma_{i_1 \dots i_p} \hat \gamma_{j_1 \dots j_q} \ve_{2}\ , \quad
  i_1 \dots i_p = 1\dots 4, 11, 12\ , \quad   j_1 \dots j_q = 5\dots 10,
\end{equation}
which is a $p$-forms in $\BR^{5,1}$ under $\SO(5,1)$ and a $q$-form in $\BR^6$ under $\SO(6)$.
Here $\bar \ve_{1}$ stands for the Dirac conjugated spinor
\begin{equation}
  \label{eq:dirac-conjugated}
  \bar \ve_1 = \ve_1^{*} \hat \gamma_{12}.
\end{equation}

The $(p,q)$ form in (\ref{eq:bilinears}) is generically nonzero if $\ve_1$ and $\ve_2$ have the same chirality
for odd $p+q$, and opposite chirality for even $p+q$. We use the spinors (\ref{eq:ve})
and (\ref{eq:so11-1-negative}) to construct the non-trivial $\SO(5,1)\times SO(6)$ $(p,q)$-forms either as 
\begin{equation}
\label{eq:Inv1}
  \rho_{i_1 \dots i_p j_1 \dots j_q} = \tilde \ve^{*} \hat \gamma_0 \hat \gamma_{i_1 \dots
    i_p} \hat \gamma_{j_1 \dots
    j_q} \ve  \quad \text{for even\ } p+q \ , 
\end{equation}
or as 
\begin{equation}
\label{eq:Inv2}
  \rho_{i_1 \dots i_p j_1 \dots j_q} =  \ve^{*} \hat \gamma_0 \hat \gamma_{i_1 \dots i_p}  
  \hat \gamma_{j_1 \dots
    j_q} \ve
  \quad \text{for odd\ } p+q \ .
\end{equation}
The forms of even degree are holomorphic in $\ve_s, \ve_c$,
while the forms of odd degree depend on $\ve_{s}, \ve_{c}$ and their complex
conjugates. Now one can easily construct a bilinear in $\rho_{p,q}$ invariants by contracting the $i$ and $j$ indexes.
We use $(a,b)$ notation to denote the standard metric pairing of the $(p,q)$-forms $a$
and $b$ as
\begin{equation}
  \label{eq:metric-pairing}
  (a,b) :=\frac 1 {p!q!} a_{i_1 \dots i_p j_1 \dots j_q} b_{i'_1 \dots i'_p j'_1 \dots j'_q} g^{i_1 i'_1} \dots g^{i_p i'_p} g^{j_1 j'_1} \dots g^{j_q j'_q}\ . 
\end{equation}
% Let also $*a$ denote Hodge dual to $a$ in $\BR^6$ or in $\BR^{5,1}$ (one
% can use dimension of $b$ to figure out the space where Hodge dual takes
% place)  and $a^{*}$ denote complex conjugate of $a$. 
Clearly, not all resulting invariants will be independent and our job will be to identify the complete set of the independent ones that parametrize the space of supercharges uniquely.
It turns out that to built independent invariants it is enough to consider (\ref{eq:metric-pairing}) with either $p$ or $q$ equal to zero. We introduce the following concise notations for the contraction of the $(p=n,q=0)$ form $\rho$ with itself
\bea
\label{eq:Invar}
I_p^n=(\rho,\rho)\ ,\quad \tilde{I}_p^n=(\rho,\rho^*)\ ,
\eea
and similarly $I_q^n,\tilde{I}_q^n$ for the invariants built out of the $(p=0,q=n)$ form. 

In the rest of this section we proceed with a systematic consideration of all cases when $\Sigma$ is non-trivial, namely $\mm = 0$ (when $\Sigma_{\BC} = \BR^{10}$)
and when $\mm \neq 0$ (when $\Sigma_{\BC} = \Sph^{6}$ or $\Sigma_{\BC} = \BR^{6}$).

%\subsection{Parametrization of supercharges when $\mm=0$}
\subsection{The case $\mm=0$,  $\Sigma_{\BC} = \BR^{10}$, $\Sigma=\BR^{4}_{\spt}$}

We start with the case when the 3-form $\mm$ (\ref{eq:form-notations}) vanishes
and the supersymmetry spinor $\ve(x)$ is pure everywhere in the space-time
$\Sigma=\BR^4_{\spt}$. In this case the pair $(\ve_s,\ve_c)$ is parametrized by $30$
real parameters where $20$ parameters parametrize a pure spinor $\ve_s$
modulo $C^{*}$ action,  and $10$ parameters $\vv^M$
define $\ve_c$ via (\ref{eq:form-notations}). Out of $30$ parameters only $2$ combinations are
invariant under the transformation of the global symmetry group. In principle
we can write down a general $30$-parameter dependent spinor $\ve$ and calculate
the invariants using (\ref{eq:metric-pairing}). But this strategy is not very practical
because in order to write the unique $\ve$ explicitly we would have to express
$30$ parameters through just two. 

It is much easier to use geometrical
intuition to cast the pair $(\ve_s,\ve_c)$ to the simplest possible form in the
first place. Let us start by choosing the simplest possible  form for a  {\it generic} pure
spinor $\ve_s$. As was discussed in section (\ref{se:susy-loops-pure-spinors}), a
pure spinor can be characterized by the complex structure $J^M_N$, or, after
lowering one index, by $10\times 10$ antisymmetric matrix $J_{MN}$. Its $4\times
4$ space-time block $J_{\mu\nu}$ can be thought of as an element in the $so(4)$
algebra. After applying an appropriate rotation of $\BR^{4}_{\spt}$, 
this $4 \times 4$ block can be transformed to a canonical form parametrized by two numbers $\alpha,\beta$
(the non-zero components can not be larger than $1$ to ensure $J^2=-1_{10\times 10}$)
\bea
J_{\mu\nu}=\left(
\begin{array}{llll}
0 & -\sin (\alpha ) & 0 & 0  \\
 \sin (\alpha ) & 0 & 0 &  \\
 0 & 0 & 0 & -\sin (\beta )  \\
 0 & 0 & \sin (\beta ) & 0 
\end{array}
\right)\ .
\label{eq:jmunu}  
\eea 
The rest of $J^M_N$ can be transformed to the canonical form below by an appropriate $SO(6)$ transformation
\begin{equation}
  \label{eq:cs-alpha-beta}
J = \left(
\begin{array}{llllllllll}
 0 & -\sin (\alpha ) & 0 & 0 & \cos (\alpha ) & 0 & 0 & 0 & 0 & 0 \\
 \sin (\alpha ) & 0 & 0 & 0 & 0 & \cos (\alpha ) & 0 & 0 & 0 & 0 \\
 0 & 0 & 0 & -\sin (\beta ) & 0 & 0 & \cos (\beta ) & 0 & 0 & 0 \\
 0 & 0 & \sin (\beta ) & 0 & 0 & 0 & 0 & \cos (\beta ) & 0 & 0 \\
 -\cos (\alpha ) & 0 & 0 & 0 & 0 & \sin (\alpha ) & 0 & 0 & 0 & 0 \\
 0 & -\cos (\alpha ) & 0 & 0 & -\sin (\alpha ) & 0 & 0 & 0 & 0 & 0 \\
 0 & 0 & -\cos (\beta ) & 0 & 0 & 0 & 0 & \sin (\beta ) & 0 & 0 \\
 0 & 0 & 0 & -\cos (\beta ) & 0 & 0 & -\sin (\beta ) & 0 & 0 & 0 \\
 0 & 0 & 0 & 0 & 0 & 0 & 0 & 0 & 0 & -1 \\
 0 & 0 & 0 & 0 & 0 & 0 & 0 & 0 & 1 & 0
\end{array}
\right)\ .
\end{equation}
To understand why this is always possible, let's take $J^M_N$ with
$J_{\mu\nu}$ given by (\ref{eq:jmunu}) and act by it on the unit vector
in the direction 1, and then choose the projection of the resulting vector on the
orthogonal compliment to 
$\BR^{4}_{\spt}$ to be the direction $5$.   
Then we do the same with the direction $2$ and call the resulting direction $6$. Notice that the directions $5$ and $6$ are orthogonal to each other because of $J^2=-1_{10\times 10}$. Similarly acting by $J$ on $3$ and $4$ gives $7$ and $8$ respectively. Eventually the remaining directions $9,10$ must transform into each other.

What we achieve at this point, using $\SO(4) \times \SO(6)$ symmetry,
 is the parametrization of the projective pure spinor $\ve_s$ by only two parameters instead of $20$ 
\begin{equation}
\label{eq:ep-alpha-beta}
\ve_{s} =  ( \cos \frac{\alpha+\beta}{2},-i  \sin \frac{\alpha-\beta}{2},0,0,i  \cos \frac{\alpha-\beta}{2}, \sin
   \frac{\alpha+\beta}{2},0,0,0,0,0,0,0,0,0,0)\ .
\end{equation}

At the next step, we reduce ten components $\vv^M$ 
parameterizing $\ve_c$ to just three components. First of all, the first four components
$\vv_1,..,\vv_4$ can be set to zero because they correspond to the special
conformal transformation of the space-time (see (\ref{eq:epsilon-special-conformal})).
As a result we are left with six parameters $\vv_5,..,\vv_{10}$. 

The projective spinor $\ve_s$ (\ref{eq:ep-alpha-beta}) is invariant under $U(1)^3$ as evident from (\ref{eq:cs-alpha-beta}). 
The first $U(1)$ simultaneously rotates the 1-2 and 5-6 planes, the second $U(1)$
simultaneously rotates the 3-4 and 7-8 planes, and the third one $U(1)$ rotates the
9-10 plane. This symmetry is enough to kill $\vv_6,\vv_8,\vv_{10}$
leaving $\vv^M=(0,0,0,0,\vv_\alpha,0,\vv_\beta,0,\vv_n,0)$. 

We summarize that the conformal supercharges $Q$ ``of the type $\mm = 0$'' can be parametrized  modulo the action of $\SO(4) \times \SO(6)$ symmetry by two angles $\alpha, \beta$  (which determine 
the complex structure $J$ at the origin $x=0$) and three non-negative real numbers $\vv_{\alpha}, \vv_{\beta}, \vv_n$. 
Still, not all five parameters $\alpha,\beta,\vv_\alpha,\vv_\beta,\vv_n$ are independent. 
Now we will use the $\SO(5,1) \times \SO(6)$ invariants (\ref{eq:Invar}) to find which points 
in the five-dimensional parametric space $(\alpha,
\beta, \vv_{\alpha}, \vv_{\beta}, \vv_{n})$ are related to each other by a $\SO(5,1)
\times \SO(6)$ transformation. 

There are just two independent  invariants (under $SO(5,1)\times SO(6)
\times \BC^{*}$)
\begin{equation}
%  \label{eq:ratio}
\begin{aligned}
\label{eq:I-1-alpha-beta-v}
I_1 &=  \frac{  \tilde{I}_p^6 } { I_p^1  }=\frac {\vv_n^2}{\vv^2} \cos ^2 \alpha   \cos ^2 \beta\ , \\
I_2 &=  \frac{  I_p^5 }  { I_p^1}=   \frac {\vv_n^2} {\vv^2}\sin^2\alpha  \sin^2 \beta  + \frac  {\vv_{\alpha }^2}{\vv^2}  \sin^2\beta  +  \frac {\vv_{\beta }^2}{ \vv^2}  \sin^2\alpha\ ,
\end{aligned}
\end{equation}
where $\vv^2 = \vv_{\alpha}^2 + \vv_{\beta}^2 + \vv_n^2$. They uniquely parametrize the conformal (projective) Killing spinor $\ve(x)$ up to the action of the global symmetry group.

Now, we would like to pick a canonical representative for the pair $(\ve_{s}, \ve_{c})$ in each $\SO(5,1) \times SO(6)$ orbit in the space of conformal killing spinors to write down concrete formulas for the Wilson loop operators. 

To choose such a representative, we analyze the allowed range of values for $I_1,I_2$ and parametrize it in a convenient way.
To find the allowed range, we fix $I_1$ and vary $I_2$ with respect to $\alpha,\beta,\vv_\alpha^2/\vv^2,\vv_\beta^2/\vv^2$. A simple calculation reveals that for the given $I_1$ the maximal value of $I_2$ is achieved when two of three terms in $I_2$ vanish. In general the invariants belong to the interval $0\leq I_1,I_2 \leq 1$ and satisfy 
\bea
\label{eq:rangei1i2}
\sqrt{I_1}+\sqrt{I_2}\leq 1\ .
\eea
The same range of allowed values could be parametrized by two parameters $\alpha, \vv^2_\beta/\vv^2$ keeping $\beta=\vv_\beta=0$. Indeed, in this case we introduce $t_1 = \vv^2_\beta/\vv^2, t_2 = \cos^2 \beta$ with $0\leq t_1,t_2 \leq 1$ and 
\begin{equation}
  \label{eq:I-1-beta-v3}
\begin{aligned}
  & I_1 =  t_1 t_2\ ,    \\
  & I_2 =  (1 - t_1)(1-t_2)\ .   
\end{aligned}
\end{equation}
Clearly $I_1,I_2$  from (\ref{eq:I-1-beta-v3}) cover the same allowed range of values (\ref{eq:rangei1i2}).

The conclusion is that for any allowed values of $I_1, I_2$ there exists
a point on the $\SO(5,1) \times \SO(6)$ orbit such that $\alpha=0,
\vv_{\beta}=0$. The space of nonequivalent pure conformal  Killing  spinors is parametrized
 by  angle $\beta$ and the cosine of the angle between the vector $v$ and the $\alpha$-plane $\vv_{\alpha}/\vv$, 
keeping $\alpha = 0$ and $\vv_{\beta} = 0$.

\subsubsection{Complex structure and Wilson loop operators}
Now we give concrete formula for the Wilson loop operator in this case. 
The most general supersymmetric coupling at point $x$ is given by $\varphi^{\bar I}Z^M_{\bar I}(x) A_M$
where $A_M=(A_\mu,\Phi_A)$, $Z^M_{\bar I}(x)$ satisfies (\ref{eq:Z-annihilate-epsilon})
and $\varphi^{\bar I}(x)$ are  five arbitrary complex numbers.
For the contour $x^{\mu}(s)$ to be a real contour in $\BR^{4}_{\spt}$ we ask $v^{\mu} = \dot x^{\mu}$ to be real.
In general the matrix $Z^M_{\bar I}(x)$ can be found with help of (\ref{eq:ZZ}). To make the connection between
$\varphi^{\bar I}$ and $dx^\mu$ more obvious it is preferable to make a transformation $Z^M_{\bar I}\rightarrow \tilde{Z}^M_{\bar I}(x)=Z^M_{\bar J}(x)U^{\bar J}_{\bar I}(x)$ bringing it to the form 
\bea
\label{eq:tZ}
\tilde{Z}^M_{\bar I}=\left(
\begin{array}{c}
I_{5\times 5}\\
-i\Theta
\end{array}
\right)\ .
\eea
The new matrix $\tilde{Z}$ is still a matrix of the antiholomorphic vectors. 
Therefore the supersymmetric coupling takes the following simple form (here $\varphi$ is an arbitrary complex number)
\bea
%\label{eq:malda-v-sus}
A_M \tilde{Z}^M_{\bar I} \left(\begin{array}{c}
\dot{x}_1\\
\dots\\
\dot{x}_4\\
\varphi
\end{array}\right)\ .
\eea

To find $\tilde{Z}(x)$ we start with the matrix $\hat Z$ at the origin which satisfies (\ref{eq:Znorm})
with $g_{MN}={1\over 2}\delta_{MN}$
and  is annihilated by $(\BI_{10\times 10}+iJ)$ with $J$ given by (\ref{eq:cs-alpha-beta})
\bea
\nonumber
(\hat Z^M_{\bar I})^T={1\over 2}
\left(
\begin{array}{llllllllll}
 \cos{\alpha \over 2}  & i \sin{\alpha \over 2} & 0 & 0 & -i\cos{\alpha \over 2} & - \sin{\alpha \over 2} & 0 & 0 & 0 & 0 \\
 -i \sin{\alpha \over 2} & \cos{\alpha \over 2} & 0 & 0 &  \sin{\alpha \over 2} & -i\cos{\alpha \over 2} & 0 & 0 & 0 & 0 \\
 0 & 0 & \cos{\beta \over 2} & i \sin{\beta \over 2} & 0 & 0 & -i \cos{\beta \over 2} & -\sin{\beta \over 2} & 0 & 0 \\
 0 & 0 & -i \sin{\beta \over 2} & \cos{\beta \over 2} & 0 & 0 & \sin{\beta \over 2} & -i \cos{\beta \over 2} & 0 & 0 \\
 0 & 0 & 0 & 0 & 0 & 0 & 0 & 0 & 1 & i
\end{array}
\right)\ .\\
\label{eq:Z1}
\eea 
The corresponding holomorphic coordinates $x^I$ on $\BR^{10}$ are
\bea
\label{eq:Xh}
\begin{array}{c}
x^{I=1}=\cos{\alpha \over 2} x_1-\sin{\alpha \over 2} x_6+i\left(\sin{\alpha \over 2} x_2-\cos{\alpha \over 2} x_5\right)\ , \\
x^{I=2}=\cos{\alpha \over 2} x_2+\sin{\alpha \over 2} x_5-i\left(\sin{\alpha \over 2} x_1+\cos{\alpha \over 2} x_6\right)\ , \\
x^{I=3}= \cos{\beta \over 2} x_3-\sin{\beta \over 2} x_8+i\left(\sin{\beta \over 2} x_4-\cos{\beta \over 2} x_7\right)\ , \\
x^{I=4}= \cos{\beta \over 2} x_4+\sin{\beta \over 2} x_7-i\left(\sin{\beta \over 2} x_3+\cos{\beta \over 2} x_8\right)\ , \\
x^{I=5}=x_9+i x_{10}\ .
\end{array}
\eea Similarly the holomorphic vector $\vv^I$ is built of $\vv^M$ with all $\vv^M=0$ except for $\vv_5=\vv_\alpha$ and $\vv_9=\vv_n$.
In fact the formulae above are too general because we can always put $\alpha=0$.
Now one can use the computer algebra to construct the $5\times 5$ matrix
$\alpha_{\bar I\bar J}$ using (\ref{eq:alpha}), calculate $Z^M_{\bar I}(x)$
using (\ref{eq:ZZ}) and then transform it to the form (\ref{eq:tZ}) by
multiplying it by an appropriate $U^{\bar J}_{\bar I}$. It is convenient to
rearrange index $M$ as follows 
\bea
A_M=(A_1,..A_4,\Phi_5,..,\Phi_{10})\rightarrow A_{\tilde M}=(A_1,..,A_4,\Phi_9,\Phi_5,..,\Phi_8,-\Phi_{10})\ .
\eea
In this case the equation (\ref{eq:ZZ}) that determines $Z$ away from the origin
obviously stays the same while the matrix (\ref{eq:Z1}) assumes a simpler form
(remember that we put angle $\alpha=0$) 
\begin{equation}
  \label{eq:matrix-Z-malda}
  \hat Z^{\tilde M}_{\bar I} = \tfrac 1 2 
   \begin{pmatrix}
     {\rm z} \\
     -i {\rm \bar z}
  \end{pmatrix}\ , \quad 
{\rm z}=\left(
\begin{array}{ccccc}
1 & 0 & 0 & 0 \\
0 & 1 & 0 & 0 \\
0 & 0 & \cos{\beta \over 2} & i \sin{\beta \over 2} & 0 \\
0 & 0 & -i \sin{\beta \over 2} & \cos{\beta \over 2} & 0 \\
0 & 0 & 0 & 0 & 1
\end{array}
\right)\ .
\end{equation}
With help of (\ref{eq:ZZ}) and using that $g_{MN}={1\over 2}\delta_{MN}$ the matrix $\Theta$ from (\ref{eq:tZ})  is given by 
\bea
\label{eq:theta-z}
\Theta=({\rm \bar z}+4{\rm z}\alpha)({\rm z}+4{\rm \bar z}\alpha)^{-1}\ ,
\eea
with $\alpha$ given by (\ref{eq:alpha}).
In general the explicit expression for $\Theta$ can be calculated with help of computer algebra.
Here we present a simple analytical derivation for the specific case $\beta=0$.
In this case $\Theta=(1-A)(1+A)^{-1}$ where matrix $A=4\alpha$ has a specific structure  $A_{ij} = a_i b_j - a_j b_i$. 
For any such matrix $A$ with arbitrary vectors  $a_i, b_i$ the inverse matrix $(1+A)^{-1}$ has a simple analytical form
\begin{equation}
  \label{eq:inverse-matrix}
  ((1+A)^{-1})_{ij} = \delta_{ij} + 
\frac{ -(a_i b_j - a_j b_i) + (ab) (a_i b_j  + b_j a_i) - a^2 b_i b_j - b^2 a_i a_j} {1  - ((ab)^2 - a^2 b^2)}\ .
\end{equation}
This immediately gives for $\Theta$
\bea
\nonumber
  \Theta^{I}_{\bar J} =  \delta^{IJ} + 2 \frac { (1 + \overline{x} \vv ) (- x^{I}  \vv^{J} +  \vv^{I} x^{J}) + (x  \vv)( x^{I} \vv^{J} +  
x^{J} \vv^{I})  - x^{I} x^{J}  \vv^2 -  \vv^I  \vv^J x^2 } { ( 1 + \overline {x} \vv)^2  - (x  \vv)^2  + x^2  \vv^2}\ , \\  
\label{eq:zeta-answer}
  x^2 \equiv x^{I} x^{I}\ , \quad  \vv^2 \equiv \vv^{I} \vv^{I}\ , \quad
 x \vv \equiv x^{I} \vv^{I}\ , \quad  \overline x \vv \equiv \overline {x^{I}} \vv^{I}\quad \quad\ .
\eea

In the generic case $\beta\neq 0$, the expression (\ref{eq:zeta-answer}) is not applicable anymore. Nevertheless, the explicit calculation reveals that 
all $\Theta^I_{\bar J}$ remain the same except for $\Theta^{A=3,4}_{\bar I}$. We
present the expressions for these couplings below and notice that they coincide with  (\ref{eq:zeta-answer}) in the limit $\beta=0$ 
\bea
\left(\Theta^{A=3}_{\bar I}\right)^T(x)={1\over \cos\beta}
\left(
\begin{array}{c}
 \frac{2 \left(i \vv_\alpha+\left(\vv_\alpha^2-\vv_n^2\right) x_1\right) \left(x_3+i \sin\beta x_4\right)}{1-2 i \vv_\alpha x_1-x^2 \left(\vv_\alpha^2-\vv_n^2\right)} \\
 \frac{2 \left(\vv_\alpha^2-\vv_n^2\right) x_2 \left(x_3+i \sin\beta x_4\right)}{1-2 i \vv_\alpha x_1-x^2 \left(\vv_\alpha^2-\vv_n^2\right)} \\
 \frac{1-2 i \vv_\alpha x_1-(\vv_\alpha^2-\vv_n^2) \left(x_1^2+x_2^2-x_3^2-2 i \sin\beta x_3 x_4+x_4^2\right)}{1-2 i \vv_\alpha x_1-x^2 \left(\vv_\alpha^2-\vv_n^2\right)} \\
 \frac{ 2   \left(\vv_\alpha^2-\vv_n^2\right) x_3 x_4+i\left(1-2 i \vv_\alpha x_1-(\vv_\alpha^2-\vv_n^2) \left(x_1^2+x_2^2+x_3^2-x_4^2\right) \right)\sin\beta}{1-2 i \vv_\alpha x_1-x^2 \left(\vv_\alpha^2-\vv_n^2\right)} \\
 -\frac{2 \vv_n \left(x_3+i \sin\beta x_4\right)}{1-2 i \vv_\alpha x_1-x^2 (\vv_\alpha^2-\vv_n^2)}
\end{array}
\right)\ ,
\eea
\bea
\left(\Theta^{A=4}_{\bar I}\right)^T(x)={1\over \cos\beta}
\left(
\begin{array}{c}
 \frac{2 \left(i \vv_\alpha+\left(\vv_\alpha^2-\vv_n^2\right) x_1\right) \left(x_4-i\sin\beta x_3 \right)}{1-2 i \vv_\alpha x_1-x^2 \left(\vv_\alpha^2-\vv_n^2\right)} \\
 \frac{2  \left(\vv_\alpha^2-\vv_n^2\right) x_2 \left(x_4-i\sin\beta x_3\right)}{1-2 i \vv_\alpha x_-x^2 \left(\vv_\alpha^2-\vv_n^2\right)} \\
 \frac{2  \left(\vv_\alpha^2-\vv_n^2\right) x_3 x_4-i\left(1-2 i \vv_\alpha x_1-(\vv_\alpha^2-\vv_n^2) \left(x_1^2+x_2^2-x_3^2+x_4^2\right)\right)\sin\beta}{1-2 i \vv_\alpha x_1-x^2 \left(\vv_\alpha^2-\vv_n^2\right)} \\
 \frac{1-2 i \vv_\alpha x_1-(\vv_\alpha^2-\vv_n^2) \left(x_1^2+x_2^2+x_3^2+2 i \sin\beta x_3 x_4-x_4^2\right)}{1-2 i \vv_\alpha x_1-x^2 \left(\vv_\alpha^2-\vv_n^2\right)} \\
 -\frac{2 \vv_n \left(x_4-i \sin\beta x_3\right)}{1-2 i \vv_\alpha x_1-x^2\left(\vv_\alpha^2-\vv_n^2\right)}
\end{array}
\right)\ .
\eea
Finally, the supersymmetric Wilson loop, parametrized by an arbitrary
contour $\gamma$ in $\BR^{4}_{\spt}$ and a complex coupling $\varphi(s)$, is
\bea
\label{eq:malda-v-sus}
W_R[\gamma(s),\varphi(s)]=\tr_{R} \Pexp \oint_\gamma  
A_{\tilde M} 
\left(
\begin{array}{c}
\BI_{5\times 5}\\
-i\Theta(x)
\end{array}
\right)
  \left(\begin{array}{c}
\dot{x}_1\\
\dots\\
\dot{x}_4\\
\varphi
\end{array}\right)
ds\ ,\\
A_{\tilde M}=(A_1,..,A_4,\Phi_9,\Phi_5,..,\Phi_8,-\Phi_{10})\ .
\eea

In the specific case $\beta=\vv_\alpha=\vv_n=0$ the operator
(\ref{eq:malda-v-sus}) 
becomes the supersymmetric Wilson loop on $\BR^{4}$ discovered by Zarembo in \cite{Zarembo:2002an}.

\subsection{The case $\mm\neq 0$, $\Sigma_{\BC} = \BR^{6}$ or $\Sigma_{\BC}=\Sph^{6}$\label{se:mneq0}}
Now we are ready to consider a more interesting case when $\mm \neq 0$, 
and hence $\Sigma$ is either a sphere $\Sph^n$ or a plane $\BR^n$ in $\BR^4$.
First of all, if $\Sigma$ is a sphere, we can always perform an appropriate
special conformal transformation 
that turns $\Sigma$ into a plane. Explicitly, such transformation amounts to
a shift of $\vv^\mu$ in such a way that $\vv_{\bar 4},\vv_{\bar 5}$ in
(\ref{eq:xiv-s6-in-basis}) vanish. Let us show that this is always possible.
We assumed that $\Sigma$ is non-trivial, hence there  are other points in $\BR^4_\spt$ besides the origin that satisfy (\ref{eq:xiv-s6-in-basis}).
Then we can choose the coordinates $x^*_\mu$ of one of those points to be the parameters of a special conformal transformation  
\bea
\vv_\mu\rightarrow \vv_\mu+{x^*_\mu\over |x^2|}\ .
\eea
Obviously such transformation kills $\vv_{\bar 4},\vv_{\bar 5}$ from (\ref{eq:xiv-s6-in-basis}). From
now on we therefore assume that $\Sigma$ is a plane $\BR^n$, $n=1,2,3,4$.
The dimension $n$ depends on the mutual orientation within
$\BR^{10}$ of the space-time $\BR^4_{\spt}$
and the pure-spinor-surface $\Sigma_{\BC} = \BR^{6}$ 

Below we classify all possible scenarios.

\subsection{$\Sigma=\BR^1$}
Perhaps the simplest scenario is when $\Sigma=\BR^{1}$.
In this case the pure spinor $\ve_s$ is unique up to a $SO(5,1)\times
SO(6)$ rotation. The main difference 
with the $\mm = 0$ case,  where $\ve_s$ was parametrized by two angles
$\alpha,\beta$, comes from the fact
 that $J$ transforms $\Sigma_{\BC}$ (and its orthogonal compliment $\Sigma_{\BC}^+$) into itself and this rigidly constraints $J$ and hence $\ve_s$. 
As always, we choose the directions $1\dots 4$  to be along the space-time
$\BR^4_{\spt} $, and we choose the direction $1$ be along
$\Sigma=\BR^1$.  Then $\Sigma_{\BC}$
includes the directions $1,5,7-10$ and its orthogonal compliment $\Sigma_{\BC}^+$ includes the directions $2-4,6$.

We can always choose the coordinate $x_5$ to be along the $J$-image of $x_1$ and $x_2$
to be along the $J$-image of $x_6$. After an appropriate $SO(4)\subset SO(6)$
rotation of the  $7-10$ directions the matrix $J$ (and the corresponding spinor
$\ve_s$) acquires the form (\ref{eq:cs-alpha-beta})  with $\alpha=0$ and $\beta={\pi/2}$.  

The corresponding complex coordinates $x^I$
are
\bea
\label{eq:R1-coord}
x^{I=1}=x_1+ix_5\ ,\\ \nonumber
x^{I=2}=x_7+i x_8\ , \\ \nonumber
x^{I=3}=x_{10}+i x_9\ , \\ \nonumber
x^{I=4}=x_2+i x_6\ , \\ \nonumber
x^{I=5}=x_4+i x_3\ .
\eea   
Thus we chose $x^{I=1,2,3}$ to lie within $\Sigma_\BC$ and $x^{I=4,5}$ to be orthogonal.

Since we assume that $\Sigma$ is $\BR^{1}$ rather than $S^1$,
$\vv_2-i\vv_6$ and $\vv_3-i\vv_4$ must vanish. We can also kill $\vv_1$
using a special conformal transformation along $\Sigma$.  

There are two independent invariants that depend on six real parameters
$\vv_5,\vv_7,..,\vv_{10}$ and one complex parameter $\mu$ from (\ref{eq:mu-3form})
\bea
\label{eq:R1-invs}
I_1 &=& -{I_q^5\over I_q^1+I_p^q}={\vv_5^2\over |\mu^2|}\ ,\\
\nonumber
I_2 &=& {I_q^5-I_q^1\over I_q^1+I_p^1}={\vv_7^2+\vv_8^2+\vv_9^2+\vv_{10}^2\over |\mu^2|}\ .
\eea 
There is a great deal of degeneracy in (\ref{eq:R1-invs}) as the two invariants depend on seven parameters. This partially can be explained by the fact that we did not fix all geometric symmetries of the setup.
There are three $U(1)$ symmetries which rotate the $7-8$, $9-10$ and both planes simultaneously. Moreover, there is dilatation that rescales all coordinates  together with $\vv^M$ and $\mu$. These symmetries allow us  to set $\mu=1$ and to kill two components out of the four $\vv_7,..,\vv_{10}$.
Another parameter can be killed because of shifts along the subspace
$\Sigma$: if we choose a different point along $\Sigma$ to be an origin
of the coordinate system then the original combination of parameters
$\vv_7,..,\vv_{10}$ will turn into a new one such that $I_1,I_2$ do not
change. Since the only invariant quantities are (\ref{eq:R1-invs}) we
can choose two variables  $\vv_5,\vv_7$ to parametrize $I_1,I_2$ while
taking $\mu_1=\Re \mu = 1,\mu_2=\Im \mu = 0$ and $\vv_M=0$ for all $M\neq 5,7$.

\subsubsection{Complex structure and supersymmetric Wilson loops}
The complex structure at the origin is given by (\ref{eq:cs-alpha-beta}) with $\alpha=0$ and $\beta=\pi/2$.
The matrix $\hat Z^M_I$ (\ref{eq:Z1}) admits the form (\ref{eq:tZ}) with $\Theta=\BI_{5\times 5}$ if we rearrange the index $M$ as follows
\bea
A_M=(A_1,..,A_4,\Phi_5,..,\Phi_{10})\rightarrow A_{\tilde M}=(A_1,\Phi_7,\Phi_{10},A_2,A_4,\Phi_8,\Phi_5,\Phi_9,\Phi_6,A_3)\ .
\eea
The components of the three-form $\mm_{\bar K \bar I \bar J}$ written in coordinates (\ref{eq:R1-coord}) are 
non-zero only if all three indexes are $1,2,$ or $3$, and zero otherwise
\begin{equation}
  \label{eq:mu-explicit}
  \mm_{\bar K \bar I \bar J} = \mu g_{\bar K K} g_{\bar I I} g_{\bar J J} \ve^{K I J},\quad \bar I,\bar J,\bar K=1,2,3\ .
\end{equation}
Here $\ve^{KIJ}$ is the absolutely antisymmetric tensor, $\ve^{123} = 1$. 
As follows from the expression for $\alpha_{\bar I\bar J}$  (\ref{eq:alpha}) and formula for $Z^M_{\bar I}$ (\ref{eq:ZZ}),
only $Z^M_{\bar I}$ for $\bar I=1,2,3$ change when we move along $\Sigma$, while $Z^M_{\bar I=4,5}$ remain the same.
Note, that one can not add the couplings 
\bea
Z^M_{\bar I=4}A_M =(A_2-i\Phi_6)\ ,\\
Z^M_{\bar I=5}A_M =(A_4-i A_3)\ ,
\eea
to the supersymmetric Wilson loop operator because
this would require the contour $\gamma$ to leave $\Sigma$. 

From now on we can neglect $Z^{\tilde M}_{\bar I}$ for $\bar I=4,5$, and assume in what follows that index $\bar I=1,2,3$.
Similarly we do not need to worry about $\tilde{M}=4,5,9,10$, and the matrix $\tilde{Z}$ effectively becomes $6\times 3$
\bea
\label{eq:Z6x3}
\tilde{Z}=\left(
\begin{array}{c}
\BI_{3\times 3}\\
-i\Theta
\end{array}
\right)\ .
\eea 

Let us define a three-dimensional vector $\tilde \alpha^{\bar K}$ dual to the 2-form $\alpha_{\bar I \bar J}$
\begin{equation}
\tilde  \alpha^{\bar K} = 2 \ve^{\bar I \bar J \bar K} \alpha_{\bar I \bar J}   = \frac { \ve^{\bar I \bar J \bar K} x^{I} \vv^{J} + \mu x^{\bar K}} { 1 + \bar x \vv}\ .
\end{equation}
The matrix $\Theta$ is then given by
\bea
 \Theta  = (1 - 4\alpha)(1 + 4\alpha)^{-1}\ , \\
  4 \alpha^{IJ} = \ve^{IJK} \alpha^{\bar K}.
\eea
This expression can be easily calculated analytically 
\bea
  \label{eq:result-for-Z}
  \Theta^{\bar I}_{ \bar J} = \frac { \delta^{\bar I \bar J} (1 - \tilde \alpha^2) + 2 \tilde \alpha^{\bar I} \tilde \alpha^{\bar J} - 2 \ve^{\bar I \bar J\bar K } \alpha^{\bar K}} { 1 + \tilde \alpha^2}, \\
   \tilde \alpha^2 \equiv \tilde \alpha^{\bar I} \tilde \alpha^{\bar I}\ . 
\eea
The resulting supersymmetric Wilson loop associated with $\Sigma$ is 
\bea
\label{eq:R1-operator}
W_R[\gamma,\varphi_1,\varphi_2]=\tr_{R} \Pexp \oint_\gamma ds
\left(
\begin{array}{c c c c c c}
A_1 & \Phi_7 & \Phi_{10} & \Phi_5 & \Phi_9 & \Phi_8
\end{array}
\right)
\left(
\begin{array}{c}
\BI_{3\times 3}\\
-i \Theta
\end{array}\right)
\left(
\begin{array}{c}
\dot{x}_1\\
\varphi_1\\
\varphi_2
\end{array}\right)
\ .
\eea
Here contour $\gamma(s)$ is just a straight line $x_1(s)$ and $\varphi_{1,2}(s)$ are arbitrary complex functions of the contour parameter $s$.
If $\varphi_1=\varphi_2=0$ the operator (\ref{eq:R1-operator})
is defined through the vector 
\bea
\Theta^{\tilde A}_1=
\left(
\begin{array}{c}
 \frac{1-2 i \vv_5 x_1+\left(1-\vv_5^2-\vv_7^2\right) x_1^2} {1-2 i \vv_5 x_1+\left(1-\vv_5^2+\vv_7^2\right) x_1^2}\\
 \frac{2 \vv_7 x_1 \left(1-i \vv_5 x_1\right)}{1-2 i \vv_5 x_1+\left(1-\vv_5^2+\vv_7^2\right) x_1^2} \\
 \frac{2 \vv_7 x_1^2}{1-2 i \vv_5 x_1+\left(1-\vv_5^2+\vv_7^2\right) x_1^2}
\end{array}
\right)\ .
\eea
If $\vv_5=0$ this Wilson loop is the conformal transformation of the
circular Wilson loop with zero expectation value from
\cite{Zarembo:2002an}. Let us notice here that the denominator $1-2 i
\vv_5 x_1+\left(1-\vv_5^2+\vv_7^2\right) x_1^2$ never vanishes and hence
the corresponding operator (\ref{eq:R1-operator}) is well defined for
any smooth $\dot\varphi_1,\dot\varphi_2$.

Besides the Wilson loops described above, there are some supersymmetric Wilson loops associated with the vector field
$u^M$ (\ref{eq:bilinear}) 
\bea
u^M \cong(0,x_3+ix_4,-x_2+ix_6,-x_6-ix_2,0,-ix_3+x_4,0,0,0,0)\ ,
\eea
living outside of $\Sigma$. 
The  components $u^\mu$ on $\BR^4_\spt$ ($x_6=\cdots =x_{10}=0$) should be real. 
Therefore $x_2=0$ and $x_3/x_4$ must be constant. The corresponding contour $\gamma:x^\mu(s)$ is a straight line
\bea
x^{\mu}(s)=(x_1,0,k_3 s, k_4 s )
\eea   
in the 3-4 plane while  $x_1$ is some constant and $x_2=0$.
The corresponding Wilson loop operator is a straight line in $\BR^4_\spt$ with the string fixed at the north pole of $S^5$ \cite{Maldacena:1998im}.

%cut here

\subsection{$\Sigma=\BR^2$}
The next scenario is $\Sigma\equiv \BR^4_\spt\bigcap \Sigma_{\BC}=\BR^2$.
In this case $J$ has the most general form (\ref{eq:cs-alpha-beta})
and  the complex coordinates on $\BR^{10}$ are 
\bea
\label{eq:Xh-R2}
\begin{array}{c}
x^{I=1}=\cos{\alpha \over 2} x_1-\sin{\beta \over 2} x_6+i\left(\sin{\alpha \over 2} x_2-\cos{\alpha \over 2} x_5\right)\ , \\
x^{I=2}=\cos{\alpha \over 2} x_2+\sin{\alpha \over 2} x_5-i\left(\sin{\alpha \over 2} x_1+\cos{\alpha \over 2} x_6\right)\ , \\
x^{I=3}=x_9+i x_{10}\ ,\\
x^{I=4}= \cos{\beta \over 2} x_3-\sin{\beta \over 2} x_8+i\left(\sin{\beta \over 2} x_4-\cos{\beta \over 2} x_7\right)\ , \\
x^{I=5}= \cos{\beta \over 2} x_4+\sin{\beta \over 2} x_7-i\left(\sin{\beta \over 2} x_3+\cos{\beta \over 2} x_8\right)\ .
\end{array}
\eea
We rearranged $x^I$ (compare with (\ref{eq:Xh})) in such a way that $x^{\mu=1,2}$ parametrize $\Sigma$ and $x^{I=1,2,3}$ parametrize $\Sigma_{\BC}$.
 
As usual,  $\vv_3,\vv_4,\vv_7,\vv_8$ vanish after a conformal transformation that makes $\Sigma$ flat, 
and we kill $\vv_1,\vv_2$ by a special conformal transformation along $\Sigma$. The only non-zero parameters are $\alpha,\beta,\mu,\vv_5,\vv_6,\vv_9,\vv_{10}$. 
There are four independent real invariants which can be combined into two real and one complex variables
\bea
\label{eq:R2-invs}
I_1&=&-{I_q^1\over I_q^1+I_p^1}{\vv_5^2+\vv_6^2+\vv_9^2+\vv_{10}^2\over |\mu^2|}\ ,\\
I_2&=&{I_q^6\over I_q^2}={((\vv_{10}-i\vv_9)\cos\alpha-\mu\sin\alpha )^2\over \mu^2}\ ,\\
I_3&=&-{\tilde{I}_p^2\over I_q^1+I_p^1}=\cos^2\beta\ .
\eea

The list of invariants is somewhat long but we still have symmetries to play with.
First of all, we can rotate the 1-2 plane  and the 5-6 plane to eliminate $\vv_6$, and the 9-10 plane  to get rid of $\Im\mu$.
Then the dilatation sets $\mu=1$, leaving  five non-trivial parameters $\alpha,\beta,\vv_5,\vv_9,\vv_{10}$.
It is not surprising that $\beta$ is an invariant. After these geometrical symmetries are used up, the only
transformation that could relate different $\beta$'s is the shift along $\Sigma\subset \Sigma_{\BC}$. Those shifts change complex structure in $\Sigma_{\BC}$,
but leave the orthogonal compliment to $\Sigma_{\BC}$ invariant. Therefore $\beta$ that governs the complex structure in $N_{\Sigma_{\BC}}$ (unlike $\alpha$ that governs the complex structure in $T_{\Sigma_{\BC}}$) is an invariant.

To completely fix the conformal gauge, we eliminate one of the four parameters $\vv_5,\vv_9,\vv_{10},\alpha$, using 
the simplified invariants:
\begin{equation}
\begin{aligned}
I_1={\vv_5^2+\vv_9^2+\vv_{10}^2}\ ,\\
\Re\ I_2^{1/2}=\cos\alpha\vv_{10}-\sin\alpha\ ,\\
\Im\ I_2^{1/2}=-\cos\alpha\vv_9\ .
\end{aligned}
\end{equation}

One easy way to do that is to express $\vv_{9}$ and $\vv_{10}$ from the last two equations and substitute into the first equation. 
We get then 
\begin{equation}
  \label{eq:v5-equation-after-subt}
   \vv_{5}^2 = I_1 -  \left ( \frac { \Im I_2^{1/2} }{ \cos \alpha} \right)^2 - \left( \frac { \Re I_2^{1/2} + \sin \alpha}
{ \cos \alpha} \right)^2
\end{equation}

For the generic values  of the invariants $I_1,I_2$ such that $\vv_5^2 > 0 $, 
the RHS of (\ref{eq:v5-equation-after-subt}) is positive. But for $\alpha$ sufficiently close to $\pi/2$
the RHS of (\ref{eq:v5-equation-after-subt}) is negative, hence it must
vanish at some intermediate value of $\alpha$. At that point $\vv_5 = 0$ and $\alpha$ can be expressed through $I_1,I_2$. In fact one can always choose $\alpha$ in such a way that $\vv_5$ vanishes leaving $\alpha, \beta, \vv_9, \vv_{10}$ as the parameters, while setting $\mu = 1, \vv_5 = \vv_6 = 0$.

\subsubsection{Complex structure and supersymmetric Wilson loops}

Similarly to the previous case $\Sigma=\BR^1$, we rearrange index $M$ as follows
\bea
A_M=(A_1,..,A_4,\Phi_5,..,\Phi_{10})\rightarrow A_{\tilde M}=(A_1,A_2,\Phi_9,A_3,A_4,\Phi_5,\Phi_6,\Phi_{10},\Phi_7,\Phi_8), \ \ \
\eea
to bring $\hat Z$ to the form (\ref{eq:matrix-Z-malda})
with ${\rm z}$ given by
\bea
\label{eq:z-alpha}
{\rm z}=\left(
\begin{array}{ccccc}
\cos{\alpha \over 2} & i \sin{\alpha \over 2} & 0 & 0 & 0\\
-i \sin{\alpha \over 2} & \cos{\alpha \over 2} & 0 & 0 & 0\\
0 & 0 & 1 & 0 & 0 \\
0 & 0 & 0 & 1 & 0 \\
0 & 0 & 0 & 0 & 1
\end{array}
\right)\ .
%\label{eq:zR2}
\eea

As in the previous case $\Sigma=\BR^1$, the two couplings $Z_{\bar I=4,5}^{\tilde M} A_{\tilde M}$ are the same for all points on $\Sigma$.
In general they can not be added to the supersymmetric Wilson loop operator because they require non-zero 
$\dot{x}^{3,4}$, and hence lead away from $\Sigma$ (in the exceptional case $\beta=\pi/2$ one of the couplings becomes $\Phi_7-i\Phi_8$ and can be added with arbitrary complex coefficient $\varphi(s)$). Therefore we neglect two last columns  ${\bar I}=4,5$ and the rows ${\tilde M}=4,5,9,10$,
similarly to the previous case, effectively reducing $Z$ to the $6\times 3$ size. The resulting matrix of the antiholomorphic vectors can be presented in the form (\ref{eq:Z6x3}) with $\Theta$ given by (\ref{eq:theta-z}) with $3\times 3$ matrix $\alpha$ given by (\ref{eq:alpha})
and the $3\times 3$ matrix %${\rm z}$
\bea
\label{eq:Z-alpha3x3}
{\rm z}=\left(
\begin{array}{ccc}
\cos{\alpha \over 2} & i \sin{\alpha \over 2}  & 0\\
-i \sin{\alpha \over 2} & \cos{\alpha \over 2}  & 0\\
0 & 0 & 1
\end{array}
\right)\ .
%\label{eq:zR2}
\eea
In a particular case, when $\alpha=0$, the matrix $\Theta$ is given by (\ref{eq:result-for-Z}).
Even in this case the explicit expression is too bulky to be written here. 

The general supersymmetric Wilson operator associated with $\Sigma$ is given by 
\bea
\label{eq:R3-operator}
W_R[\gamma,\varphi_1]=\tr_{R} \Pexp \oint_\gamma ds
\left(
\begin{array}{c c c c c c}
A_1 & A_2 & \Phi_9 & \Phi_5 & \Phi_6 & \Phi_{10}
\end{array}
\right)
\left(
\begin{array}{c}
\BI_{3\times 3}\\
-i \Theta
\end{array}\right)
\left(
\begin{array}{c}
\dot{x}_1\\
\dot{x}_2\\
\varphi_1
\end{array}\right)
\ ,
\eea
with contour $\gamma$ living on $\Sigma=\BR^2$.

In the special case $\alpha=\vv_M=0$ the matrix $\Theta$ acquires a simple form
\bea
\label{eq:Theta-Drukker}
\Theta^{\bar I \bar J}={\delta^{\bar I\bar J}(1-{\rm x}^2)+2{\rm x}^{\bar I}{\rm x}^{\bar J}-2\ve^{\bar I\bar J\bar K}{\rm x}^{\bar K} \over 1+{\rm x}^2}\ ,\\
{\rm x}^{\bar I}=(x_1,x_2,0)\ .
\eea
These loops are related by a conformal transformation to the particular case of the DGRT loops on $S^3$
\cite{Drukker:2007qr} when the contour 
is limited to the equator $S^2\subset S^3$.

The vector field $u^M$ (\ref{eq:bilinear}) 
\bea
\nonumber
u^M\cong (0, 0, \cos\beta x_4 + \sin\beta x_7 - i x_8, -\cos\beta x_3 + i x_7 + 
  \sin\beta x_8,\\ 0, 0, -\sin\beta x_3 - i x_4 - \cos\beta x_8, 
 i x_3 - \sin\beta x_4 + \cos\beta x_7, 0, 0)\ ,
\eea
gives rise to the suppersymmetric Wilson loops along the concentric
circles in the 3-4 plane for any fixed $x_1,x_2$. The corresponding
operators are the non-equator circular lines on $S^4$ with $\beta$
playing the role of the latitude \cite{Pestun:2007rz}.

\subsection{$\Sigma=\BR^3$}
In the case $\Sigma\equiv \BR^4\bigcap \Sigma_{\BC}=\BR^3$ one of the angles $\alpha,\beta$ must vanish.
We choose $\beta=0$ with the directions $1,2,3$ and $5,6,7$ to lie along $\Sigma_{\BC}$.
The corresponding complex structure is given by (\ref{eq:cs-alpha-beta}) and the holomorphic  coordinates are
\bea
\label{eq:Xh-R3}
\begin{array}{c}
x^{I=1}=\cos{\alpha \over 2} x_1-\sin{\beta \over 2} x_6+i\left(\sin{\alpha \over 2} x_2-\cos{\alpha \over 2} x_5\right)\ , \\
x^{I=2}=\cos{\alpha \over 2} x_2+\sin{\alpha \over 2} x_5-i\left(\sin{\alpha \over 2} x_1+\cos{\alpha \over 2} x_6\right)\ , \\
x^{I=3}=x_3-i x_7\ ,\\
x^{I=4}=x_4-ix_{8} , \\
x^{I=5}=x_9+ix_{10}\ ,
\end{array}
\eea
with $x_{1,2,3}$ parameterizing $\Sigma$.

As usually $\vv_1,\vv_2,\vv_3,\vv_4,\vv_8,\vv_9,\vv_{10}$ vanish after an appropriate special conformal transformation and we end up with $\mu$
and $\vv_5,\vv_6,\vv_7$. There are two invariants
\bea
\label{eq:R3-invs}
I_1&=&-{I_q^1\over I_q^1+I_p^1}={\vv_5^2+\vv_6^2+\vv_7^2\over |\mu^2|}\ ,\\
I_2&=&-{I_q^5\over I_q^1+I_p^1}={(\Re\ \mu \cos\alpha-\vv_7\sin\alpha)^2\over |\mu^2|}\ .
\eea
The $U(1)$ symmetry that rotates the 1-2 and 5-6 planes can be used to
eliminate the phase of $\mu$  and then
we use dilatation to set $\mu=1$.
It is also clear that we can always choose $\vv_6^2$ to be zero as it is always combined with $\vv^2_5$ in (\ref{eq:R3-invs}).
It is clear then that the invariant $I_1$ is an arbitrary positive number when $I_2$ is any postive number in the range 
\bea
0\leq I_2\leq 1+I_1\ .
\eea
One can cover exactly the same range by letting $\vv_5$ vanish, leaving $\vv_7$ and $\alpha$ as the only independent variables.

\subsubsection{Complex structure and supersymmetric Wilson loops}
This case is very similar to the previous one $\Sigma=\BR^2$. After rearranging index $M$
\bea
A_M=(A_1,..,A_4,\Phi_5,..,\Phi_{10})\rightarrow A_{\tilde M}=(A_1,..,A_4,\Phi_9,\Phi_5,..,\Phi_8,-\Phi_{10}).
\eea
the matrix $\hat Z$ acquires the form (\ref{eq:matrix-Z-malda}) with ${\rm z}$ given by (\ref{eq:z-alpha}).
The last two columns $Z^M_{\bar I=4,5}$ are the same everywhere on $\Sigma$. One of the corresponding couplings
$Z_{\bar i= 5}^M A_M=\Phi_9+i\Phi_{10}$ can be added to the supersymmetric Wilson operator with arbitrary complex coefficient 
although the other one $Z_{\bar I=5}^M A_M=A_4-i\Phi_8$ requires non-zero $\dot{x}_4$ and therefore leads outside of $\Sigma$. 
Upon elementating two last columns and the $4,5,9,10$ rows, the $6\times 3$ matrix $\hat Z$ becomes of the form
(\ref{eq:matrix-Z-malda}) with ${\rm z}$ given by (\ref{eq:Z-alpha3x3}).

The matrix of antiholomorphic vectors can be presented in the form (\ref{eq:Z6x3}) with $\Theta$ given by
(\ref{eq:theta-z}) with $3\times 3$ matrix $\alpha$ given by (\ref{eq:alpha})
and the $3\times 3$ matrix ${\rm z}$ (\ref{eq:Z-alpha3x3}). If $\alpha=0$ the result simplifies and $\Theta$ is given by 
(\ref{eq:result-for-Z}) but even in this case the explicit expression is too bulky to be written here. 

The general supersymmetric Wilson operator associated with $\Sigma$ is given by 
\bea
\nonumber
\label{eq:R2-operator}
W_R[\gamma,\varphi]=\tr_{R} \Pexp \oint_\gamma ds
\left(
\begin{array}{c c c c c c}
A_1 & A_2 & A_3 & \Phi_5 & \Phi_6 & \Phi_7
\end{array}
\right)
\left(
\begin{array}{c}
\BI_{3\times 3}\\
-i \Theta
\end{array}\right)
\left(
\begin{array}{c}
\dot{x}_1\\
\dot{x}_2\\
\dot{x}_3
\end{array}\right)+\\+ds \varphi (\Phi_9+i\Phi_{10})
\ ,
\eea
with contour $\gamma$ living on $\Sigma=\BR^3$.

In the special case $\alpha=\vv_M=0$ the matrix $\Theta$ acquires simple form (\ref{eq:Theta-Drukker})
with ${\rm x}^{\bar I}=(x_1,x_2,x_3)$. These loops are related by a conformal transformation to  the DGRT loops on
$S^3$~\cite{Drukker:2007qr}.

The space-time part of the vector field $u^M$ (\ref{eq:bilinear})
\bea
u^M=x_4(0,\dots,0,1,i),
\eea is zero on the boundary
$x_5=\dots=x_{10}=0$ and therefore there are no suppersymmetric Wilson
loops  besides those described above and the local operator $\Phi_9+i\Phi_{10}$.

\subsection{$\Sigma=\BR^4$}
The exotic case is when $\Sigma$ coincides with the total space-time
$\Sigma\equiv \BR^4_{\spt} \bigcap \Sigma_{\BC}=\BR^4_{\spt}$.
If we  choose the directions $5,6$ to lie inside $\Sigma$ and be defined
in the same way as in the cases $\Sigma=\BR^{2,3}$ above, the complex
structure $J$ will
 be given by (\ref{eq:cs-alpha-beta}) with some $\alpha$ and $\beta=\pi/2$.
The remaining parameters $\alpha,\mu,\vv_5,\vv_6$ form the unique invariant
\bea
\label{eq:Inv-R4}
I_1={I_q^1\over I_q^1+I_p^1}={\vv_5^2+\vv_6^2\over |\mu^2|}\ .
\eea
Clearly we can set $\mu=1$ as we did before, and also $\alpha=0$ because $I_1$ is $\alpha$-independent.
It also follows from (\ref{eq:Inv-R4}) that we can fix $\vv_6=0$ leaving $\vv_5$ to be the only non-trivial parameter. 

\subsubsection{Complex structure and supersymmetric Wilson loops}
Since we fixed $\alpha=0$ the appropriate choice of holomorphic coordinates on $\BR^{10}$
with first three coordinates $x^{I=1,2,3}$ parametrizing $\Sigma_\BC$ is
\bea
x^{I=1}=x_1+ix_5\ ,\\
x^{I=2}=x_2+ix_6\ ,\\
x^{I=3}=x_3-ix_4\ ,\\
x^{I=4}=x_7+ix_8\ ,\\
x^{I=5}=x_9-ix_{10}\ .
\eea
We rearrange index $M$
\bea
A_{M}=(A_1,..,A_4,\Phi_5,..,\Phi_{10})\rightarrow A_{\tilde
M}=(A_1,A_2,A_3,\Phi_7,\Phi_9,\Phi_5,\Phi_6,-A_4,\Phi_8,-\Phi_{10})\ , \nonumber
\eea
to bring $\hat Z$ to the form (\ref{eq:matrix-Z-malda}) with ${\rm z}=\BI_{5\times 5}$.
As in the previous cases, we remove the last two columns, which correspond to the couplings 
$\Phi_7-i\Phi_8$ and $\Phi_9+i\Phi_{10}$ (these couplings should be added to the supersymmetric Wilson loop operator) 
and the rows $4,5,9,10$ from $\hat Z$ to reduce it  (and consequently $\tilde{Z}$)  to the form (\ref{eq:Z6x3}). The matrix 
$\Theta$ is given by (\ref{eq:result-for-Z}) and the supersymmetric Wilson loop operator is
\bea
\nonumber
\label{eq:R4-operator}
W_R[\gamma,\varphi_1,\varphi_2]=\tr_{R} \Pexp \oint_\gamma ds
\left(
\begin{array}{c c c c c c}
A_1 & A_2 & A_3 & \Phi_5 & \Phi_6 & -A_4
\end{array}
\right)
\left(
\begin{array}{c}
\BI_{3\times 3}\\
-i \Theta
\end{array}\right)
\left(
\begin{array}{c}
\dot{x}_1\\
\dot{x}_2\\
\varphi
\end{array}\right)+\\
ds\varphi_1(\Phi_7-i\Phi_8)+ds\varphi_2(\Phi_9+i\Phi_{10})\ .
\eea

By definition of the Wilson loop in $\BR^4_\spt$ we need to choose contour $\gamma$ and function $\varphi$ such that the coefficients in front of all four $A_1,..,A_4$ are real for all $s$. At each point on $\Sigma=\BR^4_\spt$ only two out of four tangent directions  
would satisfy this requirement with an appropriately chosen $\varphi$. First of all, $\varphi$ must be real $\varphi=\dot{x}_3$ to avoid 
multiplying $A_3$ by a complex number. Moreover it should be such that the coefficient in front of $A_4$ is real as well (we denote it by $\dot{x}_4$)
\bea
\label{eq:goodcontour}
\sum_{\mu=1}^3 \Im \left( i\Theta^3_\mu \dot{x}_\mu \right)=0\ ,\\
\sum_{\mu=1}^3 \Re \left( i\Theta^3_\mu \dot{x}_\mu \right)=\dot{x}_4\ . \nonumber
\eea
The resulting two-dimensional vector space is quite complicated and we do not present the explicit expression for the vectors $\dot{x}^\mu(\dot{x}_1,\dot{x}_2)$  here\footnote{This vector space can be defined as a zero eigenspace of the projector matrix $\BI_{4\times 4}-P$, where the projector $P$ is a properly normalized combination $\BI_{4\times 4}+J_{4\times 4}^2$ with $J_{4\times 4}$ being the $4\times 4$ upper-left corner  part of the complex structure matrix $J^M_N(x)$.}.
We will call the contours  that satisfy (\ref{eq:goodcontour}) \emph{allowed} and from now on assume that $\gamma(s)$ is one of them. 
Since the space of allowed directions is two dimensional at each
point the contour can be parametrized by an initial point and one real
degree of freedom. We note that commutator
of 
two generic non-collinear allowed vectors 
at a given point is not an allowed vector 
\bea
\nonumber
\xi_1=\dot{x}^\mu(1,0)\ , \quad \xi_2=\dot{x}^\mu(0,1)\ ,\\
\left[\xi_1,\xi_2\right]\wedge \xi_1\wedge \xi_2\neq 0\ .
\eea
 Therefore the space of allowed directions at each point $x\in
 \Sigma=\BR^4_\spt$ can not be thought of
as a tangent-space to some two-dimensional submanifold in
$\BR^4_\spt$. Even more so,
the commutators of the commutators would span the whole four-dimensional space
\bea
[\xi_1,[\xi_1,\xi_2]]\wedge[\xi_1,\xi_2]\wedge \xi_1\wedge \xi_2\neq 0\ ,
\eea 
which means that the contour $\gamma$ is not restricted to any particular submanifold in $\BR^4_\spt$.
In this sense $\gamma$ is four-dimensional. Given that it is parametrized by only one real function (which chooses the angle on the allowed plane at each point) there is not enough degrees of freedom to ensure that $\gamma$ is closed. Therefore our general predictions would be that the contour $\gamma$ that locally ensures gauge symmetry is not closed and can not be used to construct a gauge-invariant Wilson loop.
Nevertheless there could be some particular examples of closed $\gamma$ which would be interesting to identify. 

To demonstrate that the contour $\gamma$ can have a non-trivial shape we consider a particular case $\vv_5=0$
and notice that in this case both vectors $\xi_1,\xi_2$ have no projection on fourth direction if calculated at $x_4=0$.
Therefore the contour $\gamma$ will stay at the plane $x_4=0$ if the original point belongs to it.
For such a contour the tangent vector can be described by
\bea
\label{x4zero}
\dot{x}_3=-2\frac{ \dot{x}_1(x_2-x_1x_3)-\dot{x}_2(x_1-x_2x_3)}{1-x_1^2-x_2^2+x_3^2}\ .
\eea 
Similarly such a contour will stay at $x_2=0$ if the starting point is at $x_2=0$ as follows from (\ref{x4zero}). In this case the contour will stretch in the 
$x_1-x_3$ plain and will be uniquely specified by the starting point. Let us introduce a complex coordinate $z=x_1+ix_3$. Then the contour $z(s)$ will satisfy $\dot{z}=1-z^2$ with the solution
\bea
z=\tanh(s+s_0)\ .
\eea
Here $s$ is a real parameter of the contour and $s_0$ is the complex number that specifies the starting point $x_1+ix_3=\tanh(s_0)$.
This contour interpolates between the points $(x_1=\pm 1,x_3=0)$ and the imaginary part of $s_0$ specifies the maximal value of $x_3=\tanh\Im s_0$. 

%\section{Discussion\label{se:discussion}}

\section{Acknowledgments}
We thank N.~Drukker, S.~Giombi and J.~Maldacena  for useful discussions.
A.D. would like to thank Physics Department of Harvard University for hospitality while part of this work was done.
The research of A.D. was supported by the Stanford Institute for Theoretical Physics and also by the DEO grant DE-FG02-90ER40542 and in part by the grant RFBR 07-02-00878, and the Grant for Support of Scientific Schools NSh-3035.2008.2. V.P. thanks Stanford Institute for Theoretical Physics for hospitality while part of this work was done. The research of V.P. has been partially supported by a Junior Fellowship at Harvard Society of Fellows, and grants NSh-3035.2008.2 and RFBR 07-02-00645.

\appendix

\section{Proof that $v^M \cong u^M$ is the unique solution if $u^M$ does not vanish \label{se:unique-sol}}
Here we show that $v_{M} = \lambda u_{M}$ with some complex non-zero $\lambda$ and $u^\mu$ given by (\ref{eq:bilinear}) is the unique solution to 
(\ref{eq:annih-equation}) if $u^M \neq 0$. First we notice
that it follows from (\ref{eq:triple-identity}) that $u_{M} u^{M} = 0$, i.e. $u^M$ is a light-like vector. 
In Euclidean signature it means that $u$ is necessarily complex. Let $u_{M}' = \Re\ u_{M}$ 
and $u''_{M} = \Im\ u_{M}$, so $u_M = u_{M}' + i u_{M}''$. Since $u^2  = 0$ we get $(u')^2 = (u'')^2$
and $u'_{M} u''_{M} = 0$. That is, the $u'$ and the $u''$ are two non-zero orthogonal vectors of equal length. 
The two-plane in $\BR^{10}$ spanned by $u'$ and $u''$ defines breakes $\SO(10)$ to $\SO(8) \times \SO(2)$. 
Let us make $\SO(10)$ transformation so that the basis vectors $9$ and $10$ are aligned 
with $u'$ and $u''$ respectively. 
Now we take the following representation of the ten-dimensional chiral
gamma-matrices $\Gamma_{M}$ 
\begin{equation}
  \label{eq:gamma-matrices-ten-dimensions}
\begin{aligned}
  \Gamma_{M} &=
  \begin{pmatrix}
    0 & E_{M}^{T}  \\
    E_{M} & 0
  \end{pmatrix}\ , \quad \quad M=1\dots 8\ , \\
  \Gamma_{9} &= 
  \begin{pmatrix} 
    1_{8 \times 8} & 0 \\
    0 & -1_{8 \times 8}
  \end{pmatrix}\ ,\quad
  \Gamma_{10} =
  i\begin{pmatrix}
    1_{8\times 8} & 0 \\
    0 & 1_{8 \times 8}
  \end{pmatrix}\ . 
\end{aligned}
\end{equation}

Here $E_{M}$ are the gamma-matrices for $SO(8)$.  They can be also
thought of as the $8 \times 8$ matrices representing left multiplication
in the octonion algebra (see e.g. Appendix A in \cite{Pestun:2007rz}).
These matrices
satisfy the standard anticommutation relations
\begin{equation}
  \label{eq:anti-commute-E-matrices}
  E_{M} E_{N}^{T} + E_{N} E_{M}^{T} = 2 \delta_{MN}\ .
\end{equation}
Since we have chosen direction $9$ to be aligned with $u'$ and direction $10$ to be aligned with $u''$
we get
\begin{equation}
  (\Gamma_9 + i \Gamma_{10}) \ve = 0\ .
\end{equation}
Written explicitly this means
\begin{equation}
  \begin{pmatrix}
    0 & 0 \\
    0 & 1 
  \end{pmatrix} 
  \begin{pmatrix}
    \ve^{u} \\
    \ve^{d}
  \end{pmatrix}  = 0\ ,
\end{equation}
where we represented the spinor $\ve\in S^+$ of $\SO(10)$ as $\bf{8_{s} \oplus 8_{c}}$ 
according to the breaking $\SO(8) \otimes \SO(2) \subset \SO(10)$. Clearly $\ve_d$ must vanish. 
The equation (\ref{eq:annih-equation}) then splits into two parts
\bea
(v_9+iv_{10})\ve^u &=&0\ ,\\
\label{eq:8-dim}
  v^i E_{i} \ve^{u} &=& 0\ ,\quad  i=1\dots 8\ .
\eea
Let us show now that there is no 8-dimensional vector $v^{i}$ which would solve the equation (\ref{eq:8-dim}).
First we assume that such a vector exists. Then we pick up any vector $p^i$ such that $p_i v_i \neq 0$
and multiply (\ref{eq:8-dim}) by $\ve^u E_{i}^{T} p_i$ from the left. Using (\ref{eq:anti-commute-E-matrices})
we get 
\begin{equation}
  (p_i v_i) (\ve^u)^ 2 = 0\ .
\end{equation}
Since $u^9=u'=(\ve^{u})^2 = \sum\limits_{\alpha=1}^8 \ve^{u}_{\alpha} \ve^{u}_{\alpha}$ is non-zero we get a contradiction.
Therefore $v_{M} \cong u_{M}$ is the only solution  to (\ref{eq:annih-equation}) if $u^M$ is not zero.

\section{Proof that a non-trivial pure spinor hypersurface requires zero $\ww$ and decomposable $\mm$. 
\label{se:app-decomp}}

We will get the result  in several steps.
First we show that a nontrivial $\Sigma$ requires $\ww = 0$. We start with the equation (\ref{eq:rho0rho4-rho2-condition})
\begin{equation}
  \label{eq:explicit-rho4rho0-rho2-condition}
  (1+ 2 i_x v) \wedge (\xi \wedge \mm + 2 i_x \ww) = \frac 1 2 ( \xi \wedge v + 2 i_x \mm)^2\ . 
\end{equation}
Our assumption is  that $\Sigma$ is a non-trivial smooth manifold passing through the origin. Therefore we can expand the 
(\ref{eq:explicit-rho4rho0-rho2-condition})up to the linear level in $x$ 
\begin{equation}
  \label{eq:linearize-rho4rho0-rho2-condition}
  \xi \wedge \mm + 2 i_x \ww = 0\ . 
\end{equation}
Here $\xi$ and $x$ are in the tangent space of $\Sigma$ at the origin. 

Now we multiply (\ref{eq:linearize-rho4rho0-rho2-condition}) by $\xi$ to get
\begin{equation}
  \label{eq: xi-ix-w=0}
  2 \xi \wedge i_x \ww = 0\ ,
\end{equation}
and then rewrite it as
\begin{equation}
  \label{eq:xi-ixw=0-part-1}
  2 \xi \wedge i_x \ww = 2 i_x \xi \ww - 2 i_x \xi \wedge \ww = 2(x,\xi) \ww = 0\ .
\end{equation}
Here we have used that $\ww$ is a form of the top degree and hence $\xi \wedge \ww =0$ for any $\xi$.
Now, for real $x^{\mu} \neq 0 $ we have
\begin{equation}
  \label{eq:xi-x}
  2i_x\xi \equiv 2(x,\xi)  = 2 x^{\bar I} g_{\bar I J} x^{J} = x^M x_M > 0\ , 
\end{equation}
and therefore $\ww=0$.

The analysis of the constraint (\ref{eq:rho2rho4-condition-simpler}) in section (\ref{se:pure-spinor-surface}) yielding 
$\vv\wedge \xi\wedge \mm=0$ did not assume that $\mm$ is decomposable. Using this we can rewrite the constraint (\ref{eq:rho0rho4-rho2-condition})
as (compare with (\ref{eq:main-xi-mu-2}))
\begin{equation}
  \label{eq:xi-mu-2}
\begin{split}
  (\xi + 2 (x,\xi) \vv ) \wedge \mm &= 2 (i_x \mm) \wedge (i_x \mm) \\
  &=2 i_x (\mm \wedge i_x \mm)\ .
\end{split}
\end{equation}
If we multiply both sides by $\xi$ we get zero in the left hand side % thanks to (\ref{eq:mu-xi-v=0}), 
implying
\begin{equation}
  0 = 2(x, \xi) (\mm \wedge i_x \mm) \ ,
\end{equation}
because the six form $\xi\wedge \mm \wedge i_x\mm=0$ exceeds the dimension of the space.
As a result we have 
\begin{equation}
\label{eq:mu-x-mu}
  \mm \wedge i_x \mm = 0.
\end{equation}
and (\ref{eq:xi-mu-2}) reduces to (\ref{eq:main-xi-mu-2}).

The equation (\ref{eq:main-xi-mu-2}) actually implies that $\mm$ is decomposable if $\Sigma$ is non-trivial.
To show that, we introduce an antisymmetric matrix (bi-vector) $\hat\mm$ as follows
\begin{equation}
{\hat\mm}^{i_4 i_5} = \frac 1 {3!} \ep^{i_1 i_2 i_3 i_4 i_5}\mm_{i_1 i_2 i_3}\ ,
\end{equation}
and 
reinterpret the equation (\ref{eq:main-xi-mu-2}) in a way that vector $\xi+2(x,\xi)\vv$ is a zero vector of matrix $\hat \mm$.
Clearly a non-zero antisymmetric matrix $\hat\mm$ must have at least one zero vector although there could be three ones if $\mm$ is decomposable.
The simplest way to understand it is to bring $\hat\mm$ to the canonical
form by
an appropriate  $SU(5)$ transformation 
\begin{equation}
\hat \mm= \begin{pmatrix}
    0 & 0 & 0 & 0 & 0 \\
    0 & 0 & \mu' & 0 & 0\\
    0 & -\mu' & 0 & 0 & 0\\  
    0 & 0 & 0 & 0 & \mu \\
    0 & 0 & 0 & -\mu & 0 
  \end{pmatrix} 
\end{equation}
in the new coordinate basis $z_1,.., z_5$.
If $\hat\mm$ has only one zero vector (i.e. both $\mu$ and $ \mu'$ are non-zero)
the equation (\ref{eq:main-xi-mu-2}) requires vector $\xi+2(x,\xi)\vv$ to be aligned with the direction $z_1$
while the equation (\ref{eq:mu-x-mu}) which can be rewritten as 
\bea
\hat \mm\wedge \hat \mm \wedge x=0\ ,
\eea
requires vector $x$ to have zero projection on that direction.
Therefore the contraction of $x$ and $\xi+2(x,\xi)\vv$ would give zero
\bea
0=i_x (\xi+2(x,\xi)\vv)=  (x,\xi) (1 + 2 (x,v))\ .
\eea
Hence for any non-zero $x$ on $\Sigma$ we have
\begin{equation}
  1 + 2(x,v) = 0\ .
\end{equation}
This equation does not have solutions for $x$ being arbitrary close to $0$, and therefore there could be no nontrivial $\Sigma$
passing through $x=0$. 

To resolve the contradiction we have to assume that $\mu'=0$ and therefore $\mm$ is decomposable 
\bea
\mm=\mu\ d\bar z_1\wedge d\bar z_2\wedge d\bar z_3\ .
\eea
In this case the equation (\ref{eq:main-mu-xi-v=0}) is automatically satisfied for any $x$ and the only remaining equation on $\Sigma$ is (\ref{eq:main-xi-mu-2}). 

\bibliography{lib}

\end{document}